\documentclass[prx,aps,twocolumn,floatfix,superscriptaddress,nofootinbib]{revtex4-2}
\usepackage{booktabs}
\usepackage{graphicx}
\usepackage{braket}
\usepackage{stix}
\usepackage{xcolor}

\usepackage{amsmath}
\usepackage{physics}

\definecolor{reviewblue}{RGB}{0,85,160}
\definecolor{reviewgreen}{RGB}{0,150,0}

\usepackage[hidelinks]{hyperref}
\usepackage[nameinlink,noabbrev]{cleveref}
\usepackage{tasks}
\usepackage{adjustbox} 

\makeatletter
\usepackage{dcolumn}
\usepackage{soul}
\usepackage{bm}
\usepackage{hyperref}
\usepackage{amsthm}
\usepackage{enumitem}
\usepackage{amsmath}

\usepackage{xr-hyper}

\usepackage{soul}
\usepackage[caption=false]{subfig}
\hypersetup{colorlinks,
	citecolor=blue
}

\usepackage{mathtools}
\usepackage{apptools}
\usepackage{nameref}
\AtAppendix{\counterwithin{lemma}{section}}

\makeatletter
\newcommand*{\addFileDependency}[1]{
  \typeout{(#1)}
  \@addtofilelist{#1}
  \IfFileExists{#1}{}{\typeout{No file #1.}}
}
\makeatother

\crefname{section}{Appendix}{Appendices}
\Crefname{section}{Appendix}{Appendices}

\renewcommand\thesubsection{\Alph{subsection}}

\begin{document}

\preprint{APS/123-QED}

\title{\textbf{Chiral Discrimination on Gate-Based Quantum Computers} 
}%

\author{Muhammad Arsalan Ali Akbar}
\affiliation{Department of Electrical and Computer Engineering, North Carolina State University, Raleigh, NC 27606}
\affiliation{ Department of Chemistry, North Carolina State University, Raleigh, North Carolina 27695}

\author{Sabre Kais}
\email{skais@ncsu.edu}
\affiliation{Department of Electrical and Computer Engineering, North Carolina State University, Raleigh, NC 27606}
\affiliation{ Department of Chemistry, North Carolina State University, Raleigh, North Carolina 27695}

\begin{abstract}
We present a novel approach to chiral discrimination using gate-based quantum processors, addressing a key challenge in adapting conventional control techniques using modern quantum computing. Schemes such as stimulated rapid adiabatic passage (STIRAP) and shortcuts to adiabaticity (STAP) have shown strong potential for enantiomer discrimination; their reliance on analog and continuous-time control makes them incompatible with digital gate-based quantum computing architectures. Here, we adapt these protocols for quantum computers by discretizing their Gaussian-shaped pulses through Trotterization. We simulate the chiral molecule 1,2-propanediol and experimentally validate this gate-based implementation on IBM quantum hardware. Our results demonstrate that this approach is a viable foundation for advancing chiral discrimination protocols, preparing the way for quantum-level manipulation of molecular chirality on accessible quantum architectures.

\end{abstract}

\maketitle


\section{Introduction}
\label{sec:level1}
Understanding and harnessing light–matter interactions is essential to addressing a wide range of complex challenges in modern science and engineering. Over the last two decades, scientists have developed different techniques to rigorously control its phase, amplitude, and spatial distribution, enabling the study of this interaction at the molecular level \cite{chu2007laser,loudon2000quantum,wen2004quantum,gibbon2005short,brif2010control}. For instance, these methods are crucial to understanding chirality. This intrinsic property arises when an object cannot superimpose onto its mirror image, and these mirror images are called enantiomers\cite{feringa1999absolute,bonner1991origin}. From dark clouds to the early Earth ages, nature has paraded a unique preference for certain chirality forms, selecting L-amino acids as building blocks of proteins and enzymes, while favoring D-sugars as key components of DNA and RNA\cite{hein2012origin}. In addition to biology, chirality is a universal phenomenon ranging from subatomic particles, such as neutrinos, to supra-molecular assemblies and even interstellar medium, comets, and meteorites\cite{yashima2016supramolecular,liu2015supramolecular,herbst2009complex}. This ubiquitous occurrence of chirality shows how critical it is to study its origin, implications, and applications in biological and physical sciences. The tragedy of thalidomide\cite{klausen2015we} in the late 1950s and early 1960s shows how important the precise control of chiral discrimination is in medicines, since one enantiomer of thalidomide caused severe congenital disabilities, whereas the other exhibited therapeutic effects.

Traditional separation methods, such as crystallization or distillation, did not work because the two enantiomers of chiral molecules exhibit identical physical properties, such as melting points, boiling points, and densities. As a result, their discrimination and purification remain a formidable task \cite{maier2001separation}. Researchers have proposed various spectroscopic techniques to address these challenges, employing external auxiliary fields to mediate the interaction with specific enantiomers. These techniques measure the absorption difference between L and R enantiomers by probing chiral perturbation, detect vibrational optical activity via inelastic scattering, or utilize chiral-induced spin selectivity \cite{nafie1997infrared,he2011determination,bloom2024chiral,naaman2020chiral}. However, many techniques often require concentrated samples and long acquisition times, which can be challenging to achieve in an experimental setup. 

Because chiral molecules cannot be superimposed, this asymmetry allows for unique quantum mechanical behaviors, particularly in their interaction with light. In the traditional Raman process \cite{herzberg1945infrared,shore1990coherent}, a sequence of three-state transitions leads to a final state that differs from the initial state through radiative excitation followed by spontaneous emission. However, the adiabatic passage technique enables complete population transfer between quantum states through an intermediate state while bypassing spontaneous decay. In particular, Stimulated Raman Adiabatic Passage (STIRAP) \cite{bergmann1998coherent,vitanov2001coherent,vitanov2017stimulated} involves a three-level configuration in which the broken symmetry of the chiral molecules allows microwave-driven rotational transitions (one-photon process) and Raman-like transitions via virtual intermediate states(two-photon process). In this cyclic population transfer (CPT), the STIRAP sequence $(|1\rangle \rightarrow |3\rangle \ \text{via}\ |2\rangle)$, combined with a direct $(|1\rangle \rightarrow |3\rangle)$ drive, created a chiral-selective population configuration. These two paths cancel out in the presence of inversion symmetry, but in a chiral molecule, the configuration remains open due to broken parity rules. Consequently, the triple product of the dipole moment components, which emerges during state transitions, has opposite signs for the enantiomers, enabling their discrimination \cite{li2008dynamic,kral2001cyclic}. 

The adiabatic passage (AP) is robust against parameter fluctuations and decoherence, but its adiabatic criterion requires a long evolution time. Therefore, researchers focus on developing shortcuts to adiabatic passage (STAP) \cite{guery2019shortcuts,torrontegui2013shortcuts,chen2010shortcut} that tailor the time-dependent Hamiltonian to enable faster evolution while preserving adiabaticity. Different methods have been proposed to speed up AP, such as iterative interaction pictures, transitionless quantum driving, and invariant-based inverse engineering. However, the basic idea is to add counteradiabatic terms \cite{wu2017shortcut} to the STIRAP Hamiltonian to suppress nonadiabatic transitions, and to shape the pulse time dependence so the system follows invariant eigenstates, thereby reaching the required final state. In various cases, the integration of both ideas has enabled more effective experimental implementations of STAP. 

An interesting aspect of STAP is its potential to translate into quantum gate operations, opening new possibilities for efficient population transfer in more intricate quantum systems. In recent years, STIRAP and STAP protocols have been explored and experimentally realized for transferring quantum states between transmon superconducting qubits \cite{kumar2016stimulated,hegade2021shortcuts,yin2022shortcuts}, and have also been demonstrated to speed up adiabatic transitions in three-level transmon circuits. Although STIRAP and shortcuts to adiabaticity (STAP) have been proven effective for population transfer of enantiomers in chiral systems, their implementation typically depends on analog pulse shapes and precise timing, which makes them difficult to realize on gate-based quantum computers that rely on discrete operations rather than continuous driving fields. Recent work has shown that, within a trapped-ion–based quantum simulator, a chiral system subjected to dipole interactions undergoes cyclic time evolution, accumulating a chirality-dependent non-Abelian geometric phase. The resulting phase difference serves as a robust mechanism for enantiomer discrimination. \cite{liu2025chiral}. However, preparing an initial state for a large chiral molecule using laser cooling and addressing the rotational states with a microwave pulse is still an open question. To bridge this gap, we adapt the core mechanism of these protocols onto equivalent quantum circuits, which makes them compatible with the inherent nature of current quantum devices.

In this work, we introduce a new direction for studying and controlling enantiomers at the quantum level by mapping the rotational energy levels of asymmetric top molecules onto a qubit basis. To represent the three-level Hamiltonian $H_{int}(t)$  with qubits, we consider two qubits with a basis set $\{|00\rangle,|01\rangle,|10\rangle,|11\rangle\}$. We implement the time evolution of the Gaussian $Q-$pulse in the quantum circuit using Trotterization, where the phase signatures of L and R enantiomers are encoded in the rotation gates at each step. Once the $Q$-pulse is encoded, the $P-$ and $S-$pulses are implemented by approximating their Gaussian envelopes using step-wise Trotterization. We implemented the STIRAP protocol on the state-vector simulator and two IBM quantum devices(ibm\_kingston and ibm\_fez) and benchmarked against exact diagonalization results, with high fidelity using just 20 discrete time steps. STIRAP requires long evolution times to achieve adiabatic state transfer, thereby reducing the scheme’s efficiency. We address this constraint through STAP, where engineered pulse shapes enable rapid and efficient state preparation with improved nonadiabatic population transfer. Its gate-based implementation on IBM quantum devices reproduces high-fidelity results, allowing fast and efficient chiral discrimination with a substantial time advantage over the STIRAP protocol.

\section{Stimulated Raman Adiabatic Passages (STIRAP)}
\begin{figure}[t!]
\centering
\includegraphics[scale=0.5]{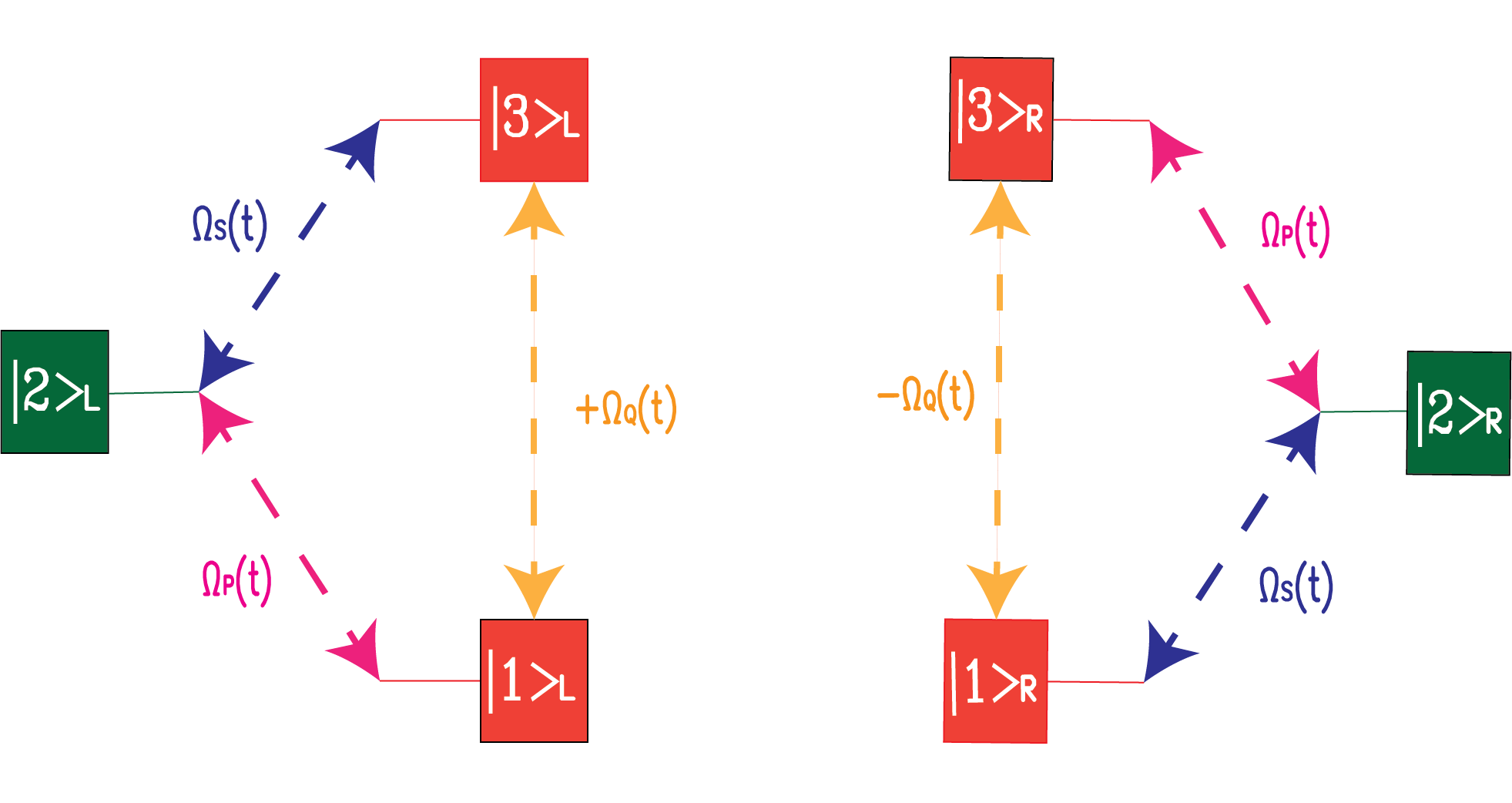}
\caption{Closed-loop coupling scheme between three discrete energy states for L and R enantiomers. The Rabi frequencies $\Omega_{P}(t)\hspace{0.3em}$, $\Omega_{S}(t)\hspace{0.3em}$, and $\Omega_{Q}(t)$ correspond to the $P-$, $S-$, and $Q-$pulses, respectively.}

\label{fig:01}
\end{figure}

Asymmetric molecules exist in two mirror-image forms, known as L and R enantiomers, as shown in Fig. \ref{fig:01}. The three rotational energy levels $|Jmk>$ for an asymmetric top are labeled $\{|1\rangle,|2\rangle,|3\rangle\}$, which are connected
in a close configuration. An external electric field drives transitions between these energy levels: $\vec{E}_{i}(t)=\hat{e}_{i}\,\alpha_{i}\cos(\omega_{i}t+\phi_{i})$
where $\alpha_{i},\hspace{0.5em}\omega_{i},\hspace{0.5em}\phi_{i},\hspace{0.5em}\text{and}\hspace{0.5em}\hat{e}_{i}$
are the amplitude, angular frequency, phase, and polarization unit vector, respectively.
The interaction between the electric field and the molecular dipole moment is given as  $\hspace{0.5em}\Omega_{ij}(t)=\vec{\mu}_{ij}\cdot\vec{E}_{i}(t)$, where $\vec{\mu}_{ij}$ denotes the dipole moment associated with the transition $ij \in \{1,2,3\}$. For both enantiomers, the electric field drives the 
$|1\rangle \leftrightarrow |2\rangle$ and $|2\rangle \leftrightarrow |3\rangle$ 
transitions with couplings $\Omega_{P}(t)$ and $\Omega_{S}(t)$ respectively. However, 
the $|1\rangle \leftrightarrow |3\rangle$ transition is characterized by $\Omega_{Q}(t) = \pm \vec{\mu}_{13} \cdot \vec{E}_{i}(t),$ where the sign difference arises from the opposite dipole moments inherent to chirality.

Then the Hamiltonian is given as

\begin{equation}
H(t)=H_{0}(t)+H_{1}(t)\label{eqn 01}
\end{equation}

where $H_{o}=\hbar\omega_{12}|2\rangle\langle2|+\hbar\omega_{13}|3\rangle\langle3|$ 
with $|1\rangle$ as the ground state (zero energy reference state). The electric field interacts with the dipole moment as \\
\begin{equation}
    H_{1}(t)=\sum_{j>i}\vec{\mu}_{ij}\cdot\vec{E}_{i}(t)
\end{equation}

where $\vec{\mu}_{12}=\vec{\mu}_{P}\cdot\vec{E}_{P}\,,\hspace{0.1cm}\vec{\mu}_{23}=\vec{\mu}_{S}\cdot\vec{E}_{S}\hspace{0.1cm}$ and $\hspace{0.1cm}\vec{\mu}_{13}=\pm\vec{\mu}_{Q}\cdot\vec{E}_{Q}$ \\
The Hamiltonian $H_{1}(t)$ becomes

\begin{equation}
\begin{split}
H_{1}(t) = \Big[ &
\vec{\mu}_{P} \cdot \vec{E}_{P} \left( |1\rangle\langle 2| + |2\rangle\langle 1| \right) \\
&+ \vec{\mu}_{S} \cdot \vec{E}_{S} \left( |2\rangle\langle 3| + |3\rangle\langle 2| \right) \\
&\pm \vec{\mu}_{Q} \cdot \vec{E}_{Q} \left( |1\rangle\langle 3| + |3\rangle\langle 1| \right)
\Big]
\label{eqn 02}
\end{split}
\end{equation}

The total Hamiltonian can be presented as a matrix with basis vectors $\{|1\rangle,|2\rangle,|3\rangle\}$

\begin{equation}
H(t)=\left(\begin{array}{ccc}
0 & \vec{\mu}_{P}\cdot\vec{E}_{P} & \pm\vec{\mu}_{Q}\cdot\vec{E}_{Q}\\
\vec{\mu}_{P}\cdot\vec{E}_{P} & \omega_{12} & \vec{\mu}_{S}\cdot\vec{E}_{S}\\
\pm\vec{\mu}_{Q}\cdot\vec{E}_{Q} & \vec{\mu}_{S}\cdot\vec{E}_{S} & \omega_{13}
\end{array}\right)\label{eqn 05}
\end{equation}

On the other hand, $\Omega_{Q}(t)$ has opposite signs for L and R enantiomers
$(\Omega_{Q}^{L}(t)=-\Omega_{Q}^{R}(t))$ for chiral molecules.

To simplify the calculations,  we transform the Hamiltonian $H(t)$
to the interaction picture and apply the Rotating Wave Approximation (RWA), assuming that the driving field is close to resonance.
The Hamiltonian in the interaction picture $H_{\text{int}}(t)$
(see Appendix \ref{app:A}  for details)

\begin{equation}
H_{Q}(t_i)=e^{iH_{0}t}H_{1}(t)e^{-iH_{0}t}\label{eqn 06}
\end{equation}
Substituting $H_{0}(t) $ and $H_{1}(t)$, the Hamiltonian matrix in the three basis becomes

\begin{equation}
H_{int}(t)=\frac{1}{2}\left(\begin{array}{ccc}
0 & \Omega_{P}e^{i\phi_{P}} & \pm\Omega_{Q}e^{i\phi_{Q}}\\
\Omega_{P}e^{-i\phi_{P}} & 0 & \Omega_{S}e^{i\phi_{S}}\\
\pm\Omega_{Q}e^{-i\phi_{Q}} & \Omega_{S}e^{-i\phi_{S}} & 0
\end{array}\right)\label{eqn 13}
\end{equation}

\section{Qubits Mapping}
\label{sec III}
\subsection{$Q-$pulse gate}

The three-level Hamiltonian $H_{\mathrm{int}}(t)$ is embedded in a two-qubit Hilbert space spanned by the basis states 
$\{|00\rangle, |01\rangle, |10\rangle, |11\rangle\}$. 
In this mapping, the logical states are assigned as 
$|1\rangle = |00\rangle$, 
$|2\rangle = |11\rangle$, 
and $|3\rangle = |10\rangle$, 
while $|01\rangle$ is designated as a leakage state that may become populated through non-adiabatic effects (see Appendix \ref{app:B}  for details).

The Hamiltonian for $Q-$pulse only is

\begin{equation}
\begin{split}
H_{int}(t)=\pm\frac{\Omega_{Q}(t)}{2}\left[e^{i\phi_{Q}}|00\rangle\langle10|+e^{-i\phi_{Q}}|10\rangle\langle00|\right]
\label{eqn 14}
\end{split}
\end{equation}

\begin{figure*}[t] 
\centering

\begin{minipage}[t]{1.0\textwidth}
    \centering
    \fbox{\includegraphics[width=\linewidth]{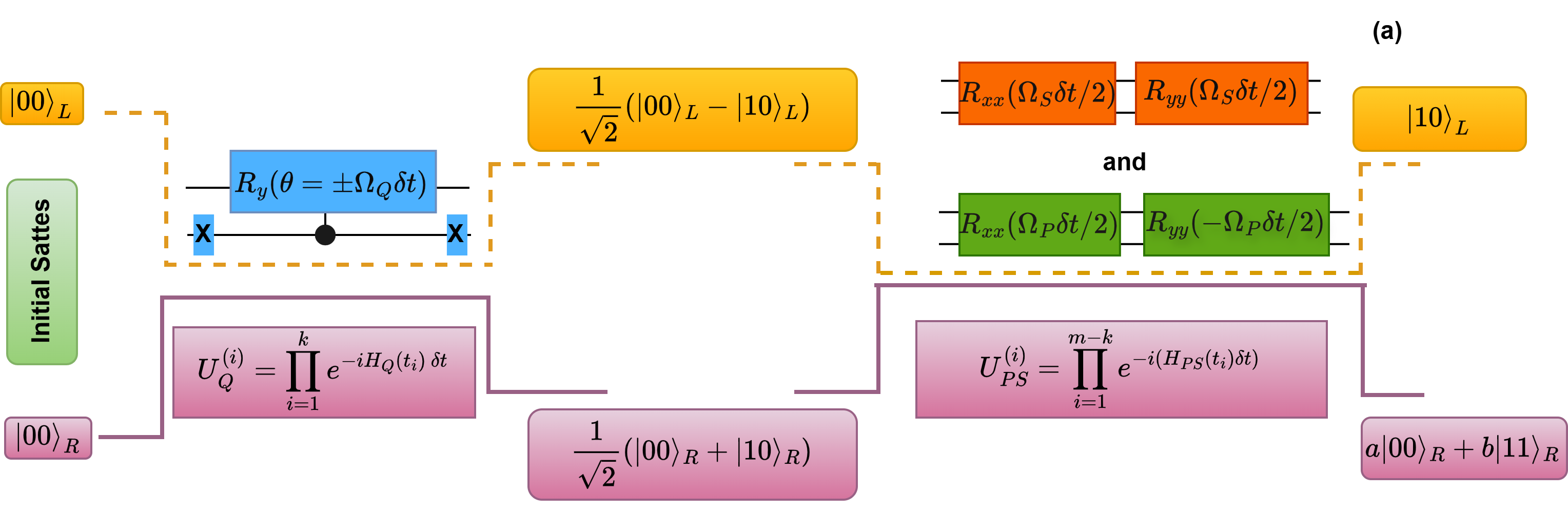}}
\end{minipage}
\vspace{0.5em} 


\begin{minipage}[t]{1.0\textwidth}
    \centering
    \fbox{\includegraphics[width=\linewidth]{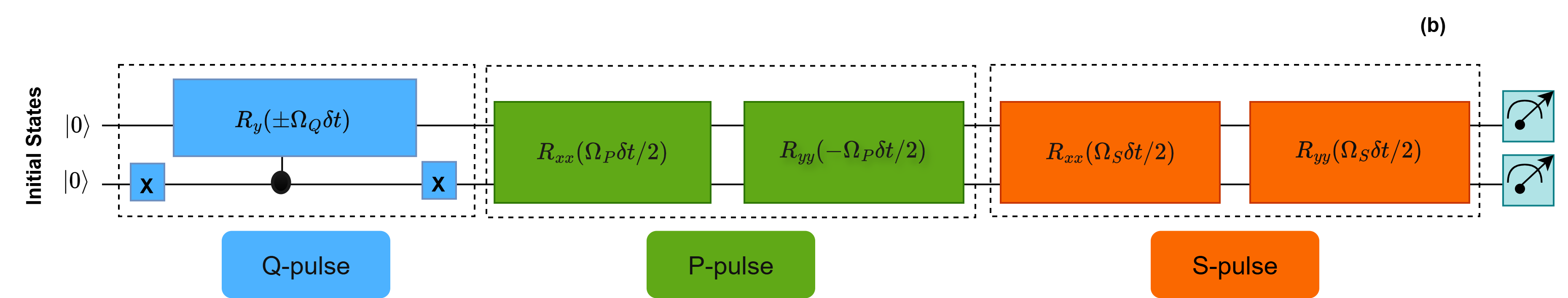}}

\caption{
\text{(a)–(b) Parametrized implementation of the Q--P--S pulse sequence realizing the proposed discrete Trotter evolution.}
Both panels depict the same logical sequence at different abstraction levels.
The protocol begins from the encoded logical state $\ket{00}_L$ (or $\ket{00}_R$), which corresponds to the physical ground state ($J=0$) for both enantiomers (Eq.\ref{Eq.35}), and evolves under the time-dependent Hamiltonians
$H_Q(t)$, $H_P(t)$, and $H_S(t)$, with each sub-step duration $\delta t$.
\textbf{(a)}~Conceptual flow diagram showing the logical evolution
$\ket{00}_L \!\rightarrow\! \frac{1}{\sqrt{2}}(\ket{00}_L - \ket{10}_L) \!\rightarrow\! \ket{10}_L$ and $\ket{00}_R \!\rightarrow\! \frac{1}{\sqrt{2}}(\ket{00}_R + \ket{10}_R) \!\rightarrow\! a\ket{00}_R+b\ket{11}_R$.
The Q-pulse (blue) applies a conditional single-qubit rotation
$R_y(\theta = \pm \Omega_Q \delta t)$ corresponding to
$U_Q^{(i)} = \prod_{i=1}^{k} e^{-iH_Q(t_i)\delta t}$.
The subsequent P-pulse (green) performs entangling operations
$R_{XX}(\Omega_P\delta t/2)$ and $R_{YY}(-\Omega_P\delta t/2)$ generated by
$U_P^{(i)} = \prod_{i=1}^{m-k} e^{-iH_P(t_i)\delta t}$,
followed by the S-pulse (orange) with
$R_{XX}(\Omega_S\delta t/2)$ and $R_{YY}(\Omega_S\delta t/2)$ governed by
$U_S^{(i)} = \prod_{i=1}^{m-k} e^{-iH_S(t_i)\delta t}$.
\textbf{(b)}~Gate-level quantum-circuit representation implementing the same
parameterized operators using native rotations and measurement qubits.
All gates are explicitly parameterized by their instantaneous Rabi frequencies
$\Omega_{Q,P,S}(t_i)$ and step time $\delta t$, ensuring full consistency with the
discrete Trotterized propagator
$U(t)\!\approx\!\prod_i e^{-iH(t_i)\delta t}$.
}

    \label{fig:02}
\end{minipage}
\end{figure*}

Here, the $Q$-pulse is defined as: $\Omega_{Q}(t) = \Omega_{0}\, e^{-(t - t_{f}/2)^{2}/T_{q}^{2}},$ where $\Omega_{0}$ and $T_{q}$ denote the peak amplitude and temporal width of the Gaussian profile, respectively. 
The phase $\phi_{Q}$ distinguishes the two enantiomers, taking the value $+\tfrac{\pi}{2}$ for the L-enantiomer and $-\tfrac{\pi}{2}$ for the R-enantiomer. The first stage starts at $t=0$ and ends at the time $t=t_{f}$ to cover the full pulse area. In the quantum circuit, the Gaussian profile of the $Q-$pulse  is simulated via Trotterized time evolution, where the Hamiltonian $H_{Q}(t_i)$ is applied in discrete intervals: $U_{Q}^{(i)} = e^{-i\big(H_{Q}(t_{i}) \, \delta t\big)}$. At each
small step $\delta t$ the instantaneous Rabi frequency $\Omega_{Q}(t_{i})$
determines the rotation angle $(\theta_{i}=\Omega_{Q}(t_{i})\delta t)$
for the $R_{y}(\theta_{i})$ gate as shown in the Fig. \ref{fig:02}. However, the quantum circuit $(CX$--$R_{y}$--$CX)$ confines the rotation to the
$\ket{00} \leftrightarrow \ket{10}$ transition, and by applying these
gates with time-dependent $\theta_{i}$, the evolution of a continuous Gaussian pulse is approximated.
 The total time
evolution for $Q-$pulse using the first-order Trotter becomes

\begin{align}
    U_{Q}^{(i)} = \prod_{i=1}^{k} e^{-i H_{Q}(t_{i}) \, \delta t}
\end{align}

\begin{figure*}[t!]
    \centering

    \begin{minipage}[t]{0.48\textwidth}
        \centering
        \includegraphics[width=\linewidth]{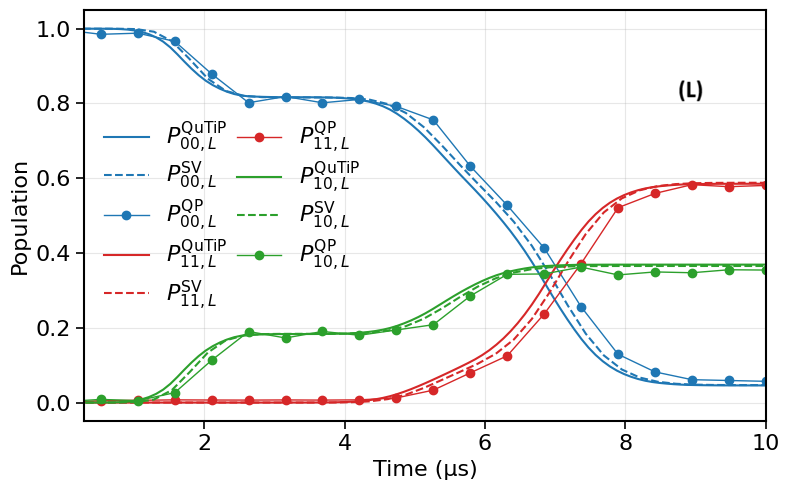}
    \end{minipage}\hfill
    \begin{minipage}[t]{0.48\textwidth}
        \centering
        \includegraphics[width=\linewidth]{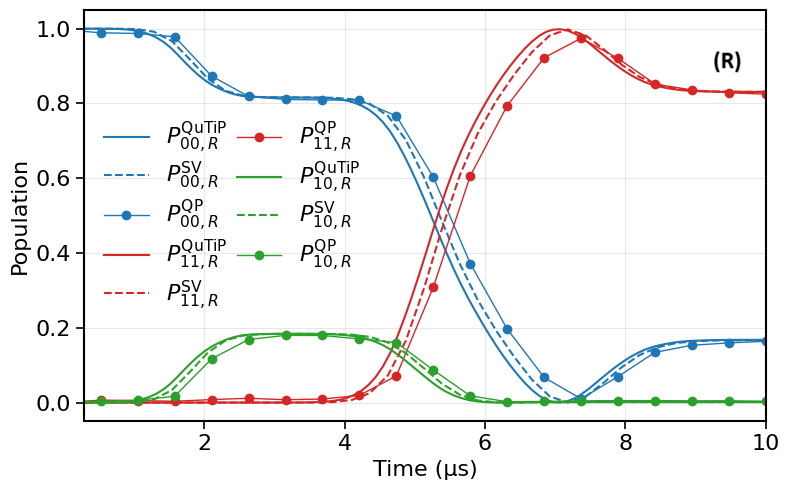}
    \end{minipage}

    \caption{Time evolution of L and R enantiomers under Stimulated Raman Adiabatic Passage (STIRAP ) protocol for $|00\rangle\hspace{0.1cm}\text{(blue)}$, $|11\rangle \text{(red)}$, and $|10\rangle \text{(green)}$ states, obtained from: (i) exact diagonalization using QuTip $(P^{(\text{QuTip})})$, (ii) statevector simulator of the discretized quantum circuit $(P^{(\text{SV})})$, and (iii) benchmarking of the exact diagonalization results with quantum processors $(P^{(\text{QP})})$.}
    \label{fig:05}
\end{figure*}

When $Q-$pulse is fully applied, the initial state of the L and R enantiomers
is driven from $|00\rangle$ to $|\Psi\rangle_{L}=\frac{1}{\sqrt{2}}(|00\rangle_{L}-|10\rangle_{L})$
and $|\Psi\rangle_{R}=\frac{1}{\sqrt{2}}(|00\rangle_{R}+|10\rangle_{R})$
respectively, as shown in Fig. \ref{fig:05}.

\subsection{$P-$ and $S-$pulse gate}

The next step is to apply the pump pulse $\Omega_{p}(t)$ and stoke
pulse $\Omega_{S}(t)$, which ultimately leads to the discrimination
between the L and R enantiomers. In the interaction picture, one gives the Hamiltonian as

\begin{equation}
\begin{split}
H_{PS}(t_i) = \frac{1}{2} \Big[ &
\Omega_{p} e^{i\phi_{p}} \, |00\rangle\langle 11| 
+ \Omega_{p} e^{-i\phi_{p}} \, |11\rangle\langle 00| \\
&+ \Omega_{S} e^{i\phi_{S}} \, |11\rangle\langle 10| 
+ \Omega_{S} e^{-i\phi_{S}} \, |10\rangle\langle 11|
\Big]
\label{eqn 17}
\end{split}
\end{equation}

Here, the resonant conditions are defined as: $\omega_{00\leftrightarrow11}-\omega_{00\leftrightarrow10}=\omega_{11\leftrightarrow10}=\omega_{S}$ and
$\omega_{00\leftrightarrow11}=\omega_{P}$. It is important to note that while the initialization Q-pulse prepares a separable state, the Hamiltonian $H_{PS}$ explicitly couples the $\ket{00}$ and $\ket{11}$ states. This interaction is non-trivial and generates entanglement. Specifically, as detailed in Sec.\ref{sec:IV}, while the L-enantiomer follows a separable dark-state trajectory ($\ket{00}\rightarrow\ket{10}$), the R-enantiomer evolves into an entangled superposition ($\alpha\ket{00}+\beta\ket{11}$). Consequently, the implementation of this protocol requires a quantum processor capable of universal entangling gates (e.g., $R_{xx}, R_{yy}$) to faithfully simulate the divergence between the separable and entangled chiral pathways.

The time-dependent pump field is modeled as a double Gaussian profile, while a single Gaussian profile represents the Stokes field.

\begin{equation}
\begin{aligned}
    \Omega_{P}(t) &= \Omega_{0} \exp\!\left[-\frac{\big(t - t_{1}) - (t_{f}-t_{1}-\tau)/2\big)^{2}}{T_{q}^{2}}\right] \\
                  &\quad + \Omega_{0} \exp\!\left[-\frac{\big(t - t_{1}) - (t_{f}-t_{1}+\tau)/2\big)^{2}}{T_{q}^{2}}\right], \\[1ex]
    \Omega_{S}(t) &= \Omega_{0} \exp\!\left[-\frac{\big(t - t_{1}) - (t_{f}-t_{1}-\tau)/2\big)^{2}}{T_{q}^{2}}\right]
\end{aligned}
\end{equation}

Here, $\Omega_{0}$ and $T_{q}$ denote the Gaussian profiles' peak amplitude and temporal width, respectively. The $\tau$ specifies the time delay between the two Gaussian components of $\Omega_{P}(t)$. The pulse shapes are not unique; what is primarily important is the adiabatic conditions $(\dot{\theta} \ll \Omega(t))$ they must satisfy between the energy levels.

The time evolution of the Gaussian profile of the $P-$ and $S-$pulse  
is encoded in the quantum circuit through Trotterization, in which
we implement $H_{PS}(t_i)$ digitally in small Trotter steps with a
discretized Gaussian pulse envelope. Applying Trotterization of
$U_{PS}^{(i)} = e^{-i\big( H_{PS}(t_{i}) \, \delta t \big)} $, the $P-$pulse collectively generates transitions between $00\leftrightarrow11$
with amplitude $\Omega_{P}(t)$ and this amplitude is encoded in the quantum  
circuit through single rotational gates $R_{x}$ with rotational
angle $\theta_{i}=\Omega_{P}(t_{i})\delta t/2$ as shown in the Fig.
\ref{fig:appB_Ppulse} and Fig. \ref{fig:appB_Spulse}.

The total time evolution for $P-$pulse and $S-$pulse using first-order Trotter as

\[
U_{PS}^{(i)}=\prod_{i=1}^{m-k}e^{-i(H_{PS}(t_{i})\delta t)}
\]

\begin{figure*}[t!] 
  \centering
  \includegraphics[width=0.48\textwidth]{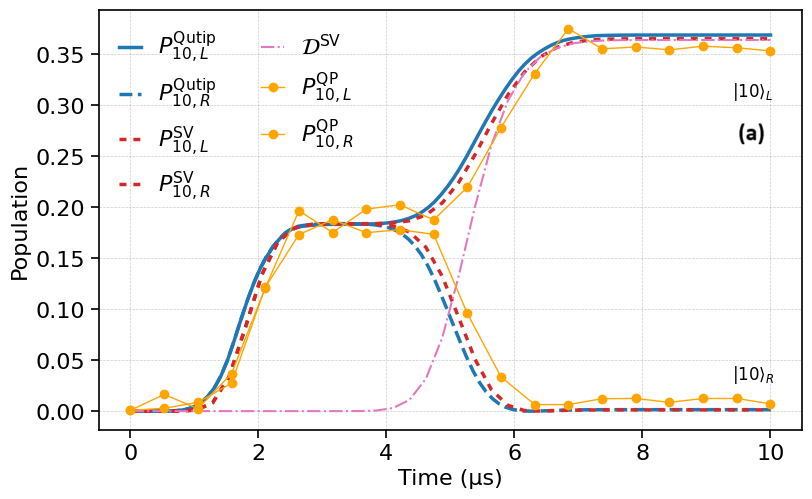}\hfill
  \includegraphics[width=0.48\textwidth]{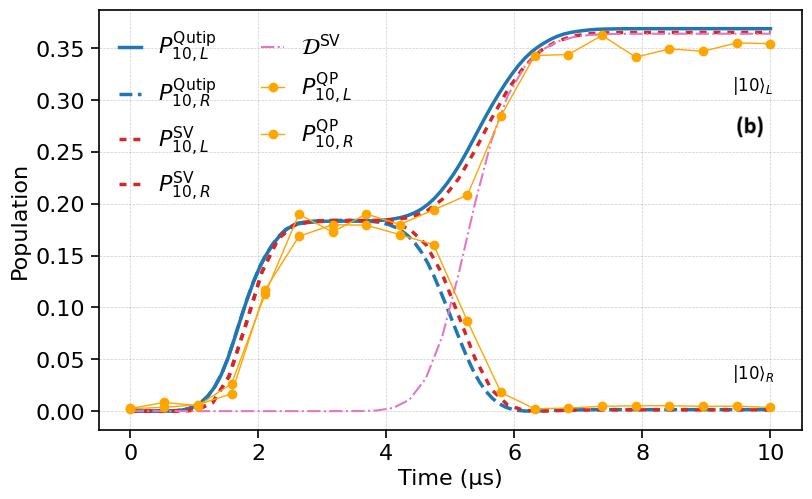}
  \caption{Chiral discrimination of L and R enantiomers using the Stimulated Raman Adiabatic Passage (STIRAP) scheme. Exact diagonalization with QuTiP (blue solid), the statevector simulator (red dashed), and quantum processor experiments (yellow solid) are shown. Panels (a) and (b) show benchmarking of the exact diagonalization results with (ibm\_kingston and ibm\_fez) quantum processors.
}
  \label{fig:04}
\end{figure*}

$k$ is the total number of steps, $m$ is the number of steps
for which the $Q-$pulse acts. At the end of the total time evolution, the
state of the L is driven from initial state $|00\rangle$ to $|\Psi\rangle_{L}=\frac{1}{\sqrt{2}}(|00\rangle_{L}-|10\rangle_{L})$
and ends up at $-|10\rangle_{L}$ as shown in the Fig. \ref{fig:04}. The matrix form for $H_{PS}(t)$ reads as

\begin{equation}
H_{PS}=\frac{1}{2}\left(\begin{array}{ccc}
0 & \Omega_{P}e^{i\phi_{P}} & 0\\
\Omega_{P}e^{-i\phi_{P}} & 0 & \Omega_{S}e^{i\phi_{S}}\\
0 & \Omega_{S}e^{-i\phi_{S}} & 0
\end{array}\right)\label{eqn 23}
\end{equation}

There are three instantaneous eigenstates with $0$ \text{and} $\pm\frac{\Omega(t)}{2}$
eigenenergies respectively.

\begin{equation}
|\gamma_{0}\rangle=\cos\alpha_{1}|00\rangle-\sin\alpha_{1}|10>\label{eqn 24}
\end{equation}

\begin{equation}
|\gamma_{\pm}\rangle=\left(e^{i(\phi_{P}+\phi_{S})}\sin\alpha_{1}|00\rangle\pm e^{i\phi_{S}}|11\rangle+\cos\alpha_{1}|10\rangle\right)\label{eqn 25}
\end{equation}

While the instantaneous eigenstates $\ket{\gamma_\pm}$ are written up to normalization for compactness; all simulated populations are computed from normalized states under Hermitian, unitary evolution.The total Rabi frequency $\Omega(t)=\sqrt{\Omega_{S}^{2}(t)+\Omega_{P}^{2}(t)}$
represents how effective the coupling strength is when the transition is
occurring between $|00\rangle\Leftrightarrow|11\rangle$ and $|11\rangle\Leftrightarrow|10\rangle$
and quantifies the energy splitting between the instantaneous eigenstates
of the Hamiltonian. When the $P$- and $S$-pulses possess nearly identical magnitudes, 
$\Omega_{S}(t)\approx\Omega_{P}(t)$, the total Rabi frequency reduces to 
$\Omega(t)\approx\sqrt{2}\,\Omega_{S}$, which reflects the equal contribution 
of each pulse. Conversely, when $\Omega_{S}(t)\approx 0$, the total Rabi frequency reduces 
to the amplitude of the $P$-pulse, indicating that the transition 
$|00\rangle \Leftrightarrow |11\rangle$ is dominated by the $P$-pulse. The mixing angle $\alpha_{1}=\tan^{-1}(\frac{\Omega_{p}}{\Omega_{S}})$ defines the system's eigenstate composition over time, leading to chiral discrimination of enantiomers. Under the
adiabatic condition $(\dot{\alpha_{1}}\ll\Omega(t))$ \cite{bergmann1998coherent,kumar2016stimulated} where $\dot{\alpha_{1}}=\frac{d}{dt}\left[\tan^{-1}(\frac{\Omega_{p}}{\Omega_{S}})\right]$ represent  
how fast the field strength is changing, ensure that the Hamiltonian
changes slowly compared to the energy gap $(\Omega(t))$ between the
eigenstate and eventually prevent the nonadiabatic coupling between
the $|\gamma_{0}\rangle$ and $|\gamma_{\pm}\rangle$, which are separated
by an energy difference of $\pm\frac{\Omega}{2}$. Conversely, if the field changes too abruptly, the time rate of change in ${\alpha}_{1}$ becomes large, and the system may undergo 
unwanted transitions between eigenstates, including leakage 
from the dark state $|\gamma_{0}\rangle$ to the bright states 
$|\gamma_{\pm}\rangle$. Consequently, start populating the intermediate state $(|2\rangle)$ rather than 
populating the target state, which is key for chiral discrimination
between two enantiomers. 

After preparing the superposition states for the two enantiomers, 
$|\Psi\rangle_{L}=\tfrac{1}{\sqrt{2}}(|00\rangle_{L}-|10\rangle_{L})$ 
and 
$|\Psi\rangle_{R}=\tfrac{1}{\sqrt{2}}(|00\rangle_{R}+|10\rangle_{R})$, 
we proceed by applying the $P$- and $S$-pulses, which are the same 
for both L and R. Utilizing fractional STIRAP \cite{vitanov1999creation,marte1991coherent}, we aim to achieve robust population transfer from initial state $|00\rangle$ to target $|10\rangle$ by adiabatically varying $\alpha_{1}(t)$ from $0$ to $\pi/2$. Initially, when $\alpha_{1}(t)\approx 0$, the dark state 
$|\gamma_{0}(t\approx 0)\rangle = \cos\alpha_{1}|00\rangle - \sin\alpha_{1}|10\rangle \approx |00\rangle$ 
is perfectly aligned with the ground state. Later, at $\alpha_{1}(t)\approx \tfrac{\pi}{4}$, 
the driving fields satisfy $\Omega_{S}(t)\approx\Omega_{P}(t)$, and the instantaneous 
eigenstate becomes 
$|\gamma_{0}(t=t_{1/2})\rangle\approx \tfrac{1}{\sqrt{2}}(|00\rangle - |10\rangle)$, 
which corresponds to maximal coherence. At the final stage, when $\alpha_{1}(t)\approx \tfrac{\pi}{2}$, the driving fields satisfy 
$\Omega_{S}(t)\ll \Omega_{P}(t)$, and the dark state evolves into 
$|\gamma_{0}(t=t_{1/2})\rangle\approx -|10\rangle$, signifying the complete transfer of population 
to the target state. Under adiabatic evolution, the L enantiomer follows the instantaneous dark 
eigenstate $|\gamma_{0}(t)\rangle$ and ultimately evolves into the target state $|10\rangle$. 
In contrast, the R-enantiomer, being orthogonal to the dark state $|\gamma_{0}(t)\rangle$,  
evolves as a superposition of the bright states $|\gamma_{\pm}(t)\rangle$.
 However, when a state evolves under a time-dependent Hamiltonian 
$H_{PS}(t)$, it acquires an overall dynamic phase,  $\Phi_{\text{dyn}}(t) \;=\; -\frac{1}{\hbar} \int E(t)\,dt$
with $E(t)$ denoting the instantaneous eigenvalue of the system 
during evolution. For a given eigenstate
$|\gamma_{\pm}(t)\rangle$ the instantaneous eigenvalue is $E_{\pm}(t)=\pm\frac{\Omega(t)}{2}$
thus the phase accumulated by $|\gamma_{\pm}(t)\rangle$ is: $\Phi_{\pm}=-\frac{1}{2}\int_{t=0}^{t=t_{f}}\Omega(t)dt$.
Eventually, the final state becomes

\begin{equation}
\begin{split}
|\Psi_{R}(t = t_{f})\rangle = \frac{1}{\sqrt{2}} \Big[ &
e^{-\frac{i}{2} \left( \int \Omega(t) \, dt \right)} 
\, |\gamma_{+}(t = t_{i})\rangle \\
&+ e^{+\frac{i}{2} \left( \int \Omega(t) \, dt \right)} 
\, |\gamma_{-}(t = t_{i})\rangle
\Big]
\label{eqn 26}
\end{split}
\end{equation}

As the applied field $\Omega_{S}(t)$ and
$\Omega_{P}(t)$ start at $t=t_{i}$ when $\alpha_{1}(t)\approx\frac{\pi}{4}$
the initial state of $\gamma_{\pm}$ becomes
$|\gamma_{\pm}(t=t_{i})\rangle=\frac{1}{\sqrt{2}}\left(|00\rangle\pm|01\rangle+|10\rangle\right)$.
Putting the value of $|\gamma_{\pm}(t=t_{i})\rangle$ in Eq. (\ref{eqn 26})
\begin{figure}[t!] 
    \centering
    
    \includegraphics[width=\linewidth, height=6cm, keepaspectratio]{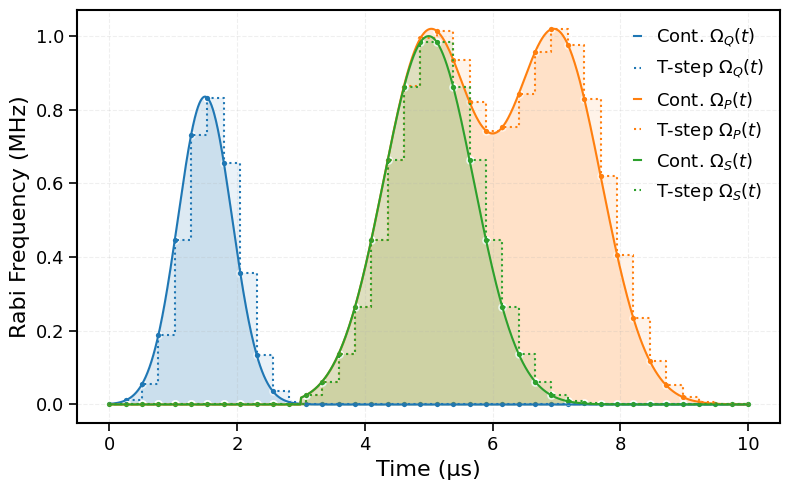}
    \caption{Solid lines show the continuous Gaussian envelopes: $\Omega_{Q}(t)$ (blue), $\Omega_{P}(t)$ (orange), and $\Omega_{S}(t)$ (green). In contrast, dashed lines depict their Trotter decompositions into steps of fixed duration $\delta t$, preserving the integrated pulse area.}
    \label{fig:03}
    
\end{figure}

\begin{equation}
\begin{split}
|\Psi_{R}(t=t_{f})\rangle 
= \frac{1}{2} \Big[ &
\cos\!\left( \tfrac{1}{2} \!\int \Omega(t)\, dt \right) 
\left( |00\rangle + |10\rangle \right) \\
& - i\,\sin\!\left( \tfrac{1}{2} \!\int \Omega(t)\, dt \right) |01\rangle
\Big]
\label{eqn 27}
\end{split}
\end{equation}

At the final time $t = t_{f}$, the mixing angle approaches 
$\alpha_{1}(t) \approx \tfrac{\pi}{2}$, corresponding to the regime 
$\Omega_{S}(t) \ll \Omega_{P}(t)$. As a result, the state of the 
R enantiomer becomes
\[
    |\Psi_{R}(t_{f})\rangle 
    = \cos\rho\,|00\rangle + \sin\rho\,|11\rangle ,
\]
with $\rho = \tfrac{1}{2}\int \Omega(t)\, dt$. In this configuration, 
the dynamics dominantly couple the states 
$|00\rangle \leftrightarrow |11\rangle$, while the population of the R enantiomer in the state
$|10\rangle$ vanishes due to destructive interference.

\begin{figure*}[t!]
  \centering

\end{figure*}

\section{Results for Stimulated Raman Adiabatic Passages (STIRAP)}
\label{sec:IV}

We use Trotter decomposition to numerically evolve the states under the 
time-dependent Hamiltonian: $\hat{H}(t)=\hat{H}_{Q}(t)+\hat{H}_{PS}(t)$. A first-order Trotter--Suzuki approximation is then applied, resulting in

\begin{equation}
e^{-i\hat{Ht}}\approx\left(e^{-i\hat{H}\delta t}\right)^{m}\approx\left(e^{-i\left\{ \hat{H}_{Q}+\hat{H}_{PS}\right\} \delta t}\right)^{m}\label{eqn 28}
\end{equation}

where $\delta t=\frac{t}{m}$ is the step size and $m$ is the total
number of steps. To generate the initial superposition through the $\Omega_{Q}(t)$ pulse, 
we first apply $k$ number of Trotter steps $(U_{Q}^{(i)} = \prod_{i=1}^{m} e^{-i H_{Q}(t_{i}) \, \delta t})$. 
Following this initialization, the subsequent evolution is governed by 
$e^{-i \hat{H}_{PS}\,\delta t}$. Therefore, the overall Trotter 
decomposition can be expressed as

\begin{equation}
e^{-i\hat{Ht}}\approx\left(e^{-i\hat{H}_{Q}\delta t}\right)^{k}\left(e^{-i\hat{H}_{PS}\delta t}\right)^{m-k}+O(\delta t^{2})\label{eqn 29}
\end{equation}

\noindent Finally, we implemented $\hspace{0.3em}\Omega_{Q}(t)\hspace{0.3em},$ $\Omega_{P}(t)\hspace{0.3em} \text{and} \hspace{0.3em}\Omega_{S}(t)$
pulses using the Trotter decomposition and divided the continuous Gaussian
pulses into a discrete rectangular shape as shown in Fig. \ref{fig:03}. The $\Omega_{Q}(t)$ pulse (blue dashed line) starts at $t=0$ and ends at 
$t \approx 3\,\mu\text{s}$. During this interval, the L and R enantiomers 
are driven from the initial state $|00\rangle$ to $|\Psi\rangle_{L}=\tfrac{1}{\sqrt{2}}(|00\rangle_{L}-|10\rangle_{L})$ 
and 
$|\Psi\rangle_{R}=\tfrac{1}{\sqrt{2}}(|00\rangle_{R}+|10\rangle_{R})$, respectively.
 The green continuous Gaussian shape represents the $\Omega_{s}(t)$ pulse, which drives the transition from the intermediate state $| 11\rangle$ to the target state $| 10\rangle$. We apply it before the $\Omega_{p}(t)$ (orange dashed line), which is responsible for $| 00\rangle$ to $| 11\rangle$ population transitions. The reason for applying $\Omega_{S}(t)$ first is intuitive; the dark state $|\gamma_{0}(t)\rangle$ is a superposition of 
$|00\rangle$ and $|10\rangle$ that decouples from $|11\rangle$ 
and is immune to the spontaneous emission of $|11\rangle$. 
Therefore, when the $\Omega_{S}(t)$ pulse is applied first, it 
couples $|11\rangle \Longleftrightarrow |10\rangle$, while the 
transition $|00\rangle \Longleftrightarrow |11\rangle$ remains 
off ($\Omega_{P}(t)=0$). In this case, the mixing angle satisfies 
$\alpha_{1}(t)=0$, and the system begins in $|00\rangle$, which 
coincides with the dark state.
 As $\Omega_{p}(t)$ starts to increase  
gradually, the mixing angle starts from $\alpha_{1}(t=0)\approx0$ to
reach an intermediate value of $\alpha_{1}(t_{1/2})\approx\frac{\pi}{4}$
and ends up to $\alpha_{1}(t=t_{f})\approx\frac{\pi}{2}$. As a result, when the applied field varies sufficiently slowly, 
the system remains in the dark state throughout the process, 
enabling population transfer from the initial state $|00\rangle$ 
to the target state $|10\rangle$ without populating the 
intermediate state $|11\rangle$.
Fig.~\ref{fig:03} illustrates that discretizing the Gaussian pulses into only 
20 Trotter steps introduces a noticeable Trotter error. Increasing the resolution 
by employing a finer grid for the pulse discretization progressively reduces this 
error, $\epsilon(1/N)$. However, the computational cost of the simulation grows 
linearly as $O(N)$, where $N$ is the number of Trotter steps.

As shown in Fig. \ref{fig:04}(a), we apply the STIRAP scheme using exact diagonalization with QuTiP (green solid line), a statevector simulator (green dashed line), and finally, we benchmark the STIRAP protocol using two IBM quantum processors (yellow dotted line).
In Fig. \ref{fig:04}(a), we employ ibm\_kingston, which exhibits results closely
matching the QuTiP exact simulation, while Fig. \ref{fig:04}(b) presents results
from ibm\_fez, where noticeable oscillations appear near the end,
likely due to hardware noise and unwanted transitions arising from
nonadiabatic effects.
The circuit was executed on both quantum processors with 5,000 shots to obtain accurate output statistics. We used the IBM Quantum Runtime Sampler, which lacks built-in error mitigation. However, no correction or suppression techniques were applied, such as gate twirling or dynamical decoupling. Sec. ~\ref {sec 01} uses the same experimental conditions.

From Fig.~\ref{fig:04}(a) and Fig.~\ref{fig:04}(b), the results of the quantum computer 
closely match the QuTiP simulations, even for $t \approx 5\,\mu s$. 
This agreement demonstrates that the system reliably evolves toward their target states once discrimination between the L and R 
enantiomers is achieved. The R enantiomer does not populate the state $|10\rangle$, as its evolution 
leads to a superposition of $|00\rangle$ and $|11\rangle$.
 Contrarily, the L molecule should be completely populated in state $|10\rangle$, but from
the results, the probability of being in state $|10\rangle$ is
less than $40\%$. The primary reason is the slow adiabatic evolution $(\dot{\alpha_{1}}\lll\Omega)$
to keep the L enantiomers in the dark state $|\gamma_{0}(t)\rangle$, but the rate of change of the pulses is higher $(\dot{\alpha_{1}}\sim\Omega)$; therefore, the system leaks into other states $(|\gamma_{\pm}\rangle)$.

Another reason for the lower fidelity of L enantiomers is imperfect
superposition created by $Q-$pulse, ideally the state
$|00\rangle$ and $|10\rangle$ should have equal superposition. In
Fig. \ref{fig:05} the green curve $(|10\rangle)$ illustrates how the L and
R enantiomers evolve toward their respective target states. 
Conversely, the red and blue curves represent the time dynamics of the
$|00\rangle$ and $|11\rangle$ states respectively. However, the measured probability of 
$|00\rangle$ and $|10\rangle$ states is lower than the theoretically 
predicted value after the $Q-$pulse is applied.

\begin{figure*}[t!]
    \centering

    \begin{minipage}[t]{0.48\textwidth}
        \centering
        \includegraphics[width=\linewidth]{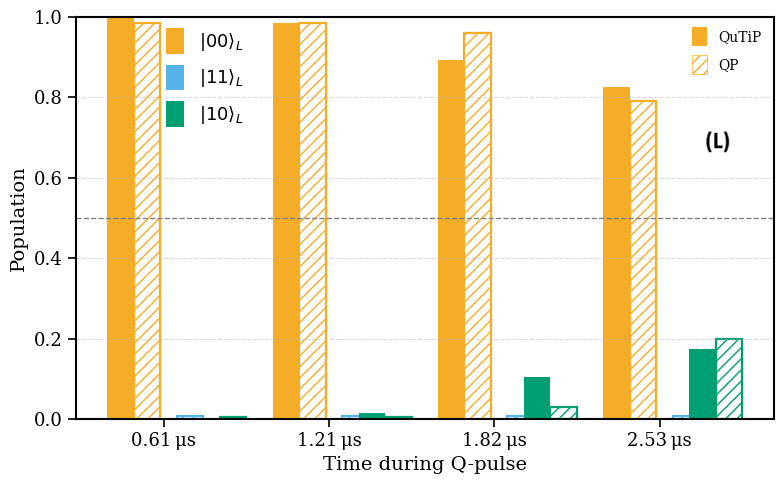}
    \end{minipage}\hfill
    \begin{minipage}[t]{0.48\textwidth}
        \centering
        \includegraphics[width=\linewidth]{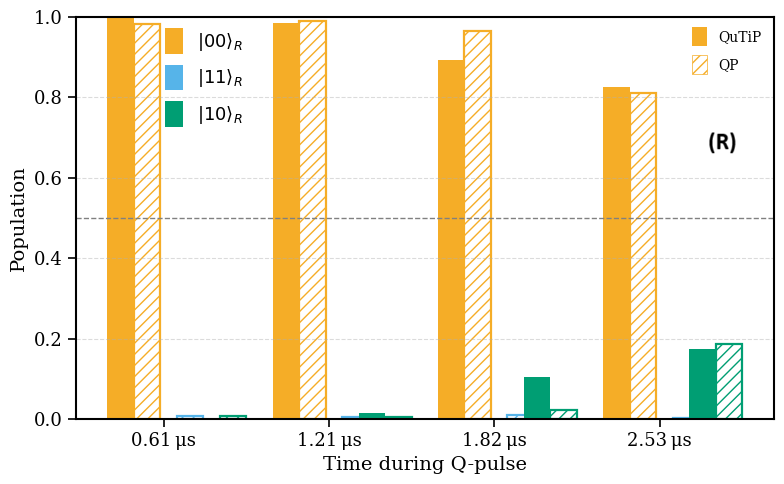}
    \end{minipage}

    \caption{State populations $|00\rangle$ (orange), $|11\rangle$ (blue), and $|10\rangle$ (green) at the midpoint ($t=0.61\,\mu s$) and end ($t=2.53\,\mu s$) of the $\Omega_{Q}(t)$ pulse, under Stimulated Raman Adiabatic Passage (STRIP) scheme. Bars with hatched fill represent results from the quantum processor, while solid bars show exact diagonalization using QuTiP, for both enantiomers.}
    \label{fig:06a}
\end{figure*}

In Fig.~\ref{fig:06a}, we present bar charts showing the state 
probabilities for $|00\rangle$, $|11\rangle$, and $|10\rangle$. 
These are evaluated at the midpoint of the $Q-$pulse 
($t = 0.61\,\mu s$) and at its end ($t = 2.53\,\mu s$). 
At $t = 2.53\,\mu s$, the measured probabilities for $|00\rangle$ 
and $|10\rangle$ are $0.82$ and $0.18$, respectively, for both 
L and R enantiomers. Ideally, the final state should 
approach an equal superposition, $|\Psi\rangle_{L}=\tfrac{1}{\sqrt{2}}(|00\rangle_{L}-|10\rangle_{L})$ 
and 
$|\Psi\rangle_{R}=\tfrac{1}{\sqrt{2}}(|00\rangle_{R}+|10\rangle_{R})$,
when the $Q-$pulse is fully applied.

\section{Shortcuts to Adiabatic Passage (STAP)}
\label{sec 01}
\subsection{Mathematical formulation}

Based on the given results, it is evident that STIRAP is not following
the adiabatic evolution $(\dot{\alpha_{1}}\lll\Omega)$ completely
which makes it less efficient to discriminate L and R enantiomers. Therefore
we need to engineer the pulse shape and instantaneous eigenstate
to get a higher population of L in state $|10\rangle$. To address this
Limitation: Shortcuts-to-Adiabatic Passage (STAP) is a  powerful technique \cite{wu2017shortcut}, 
in which we tailored controlled pulses that enable state transfer
into the target state in a short interval of time by suppressing unwanted
transitions. 

Consider the same three-level system $(|00\rangle,|11\rangle,\text{and}  \hspace{0.1em}|10\rangle)$
interacting with a time-dependent controlled field $\Omega_{P}(t)$
and $\Omega_{S}(t)$. The Hamiltonian in the Rotating 
Wave Approximation (RWA) frame becomes

\begin{equation}
H_{PS}(t)=\Omega_{P}(t)|00\rangle\langle11|+\Omega_{S}(t)|11\rangle\langle10|+\text{h.c.}\label{eqn 30}
\end{equation}

The three instantaneous eigenvalues $0$ and $\pm\frac{\Omega(t)}{2}$ corresponding to eigenstates are

\begin{equation}
|\gamma_{0}\rangle=\cos\alpha_{1}|00\rangle-\sin\alpha_{1}|10>\label{eqn 31}
\end{equation}

\begin{equation}
|\gamma_{\pm}\rangle=\left(e^{i(\phi_{P}+\phi_{S})}\sin\alpha_{1}|00\rangle\pm e^{i\phi_{S}}|11\rangle+\cos\alpha_{1}|10\rangle\right)\label{eqn 32}
\end{equation}

The transformation of $H_{PS}(t)$ into
adiabatic frame gives two main results: (i) instantaneous eigenstate
$(\gamma_{\pm,0}(t)\rangle)$ transforms into time independent states  
and non-adiabatic coupling terms $(\dot{\alpha_{1}}(t))$ appear explicitly
in the transformed Hamiltonian $H_{I}(t)$. The detailed derivation is in the Appendix \ref{app:C}. The adiabatic transformation
is given as

\begin{equation}
U_{0}(t)=\sum_{n=0,\pm}|\gamma_{n}\rangle\langle\gamma_{n}(t)|\label{eqn 33a}
\end{equation}

The Hamiltonian in the adiabatic frame takes the form

\begin{equation}
H_{I}(t)=U_{0}(t)H_{PS}(t)U_{0}^{\dagger}(t)-iU_{0}(t)\dot{U}_{0}^{\dagger}(t)\label{eqn 34a}
\end{equation}

Putting the value of $U_{0}(t)$ in Eq. (\ref{eqn 34a}) 

\begin{equation}
\begin{split}
H_{I}(t) = &\ \Omega(t) 
\Big[ |\gamma_{+}\rangle\langle\gamma_{+}| - |\gamma_{-}\rangle\langle\gamma_{-}| \Big] \\
&+ \frac{\dot{\alpha}_{1}(t)}{\sqrt{2}} 
\Big[ |\gamma_{+}\rangle\langle\gamma_{0}| + |\gamma_{-}\rangle\langle\gamma_{0}| \Big] 
+ \text{h.c.}
\label{eqn 35}
\end{split}
\end{equation}

The nonadiabatic coupling terms, which are now directly visible in the adiabatic
frame, are responsible for the leakage of dark state $|\gamma_{0}(t)\rangle$ into  
$|\gamma_{\pm}(t)\rangle$. Now the goal of STAP is to cancel these
nonadiabatic coupling terms by introducing a controlled auxiliary field
$\Omega_{P}^{'}(t)$ and $\Omega_{S}^{'}(t)$, and this was not possible
without transforming the Hamiltonian $H_{PS}(t)$ into the adiabatic
frame. Designing the auxiliary pulses is difficult because the unwanted coupling is hidden in the lab frame.

Combining the controlled auxiliary field with the STIRAP protocol speeds up the population from the initial state to the target state. In this case, the Hamiltonian takes the form

\begin{equation}
\begin{split}
H'(t) = &\ \big[ \Omega_{P}(t) + \Omega_{P}'(t) \big] |00\rangle\langle 11| \\
&+ \big[ \Omega_{S}(t) + \Omega_{S}'(t) \big] |11\rangle\langle 10| 
+ \text{h.c.}
\label{eqn 36}
\end{split}
\end{equation}

where $\Omega_{P}^{'}(t)$ and $\Omega_{S}^{'}(t)$ are the auxiliary
pulses combined with original pulses $\Omega_{P}(t)\hspace{0.3em} \text{and}\hspace{0.3em}\Omega_{S}(t)$ respectively.
For $|\varphi_{0}(t)\rangle$ to act as an evolution state for modifying  
Hamiltonian $H^{'}(t)$, it must guarantee the same final state $|\gamma_{0}(t)\rangle$.
Therefore, we can choose the following dressed state, which ensures the same
final state

\begin{equation}
\begin{split}
|\varphi_{0}(t)\rangle = \cos\theta_{2}(t) \Big[ &
\cos\alpha_{1}(t) \, |00\rangle 
+ e^{i\phi} \sin\alpha_{2}(t) \, |11\rangle \\
&+ \sin\alpha_{1}(t) \, |10\rangle
\Big]
\label{eqn 37}
\end{split}
\end{equation}

The following boundary conditions  will ensure the same final state $|\gamma_{0}(t)\rangle$ 

\begin{equation}
\left\{
\begin{aligned}
   \alpha_{1}(t=0) &= \tfrac{\pi}{4}, \\
   \alpha_{1}(t=t_{f}) &= \tfrac{\pi}{2}, \\
   \alpha_{2}(t=0) &= \alpha_{2}(t=t_{f}) = 0
\end{aligned}
\right.
\end{equation}

The other dressed states $|\varphi{}_{\pm}(t)\rangle$ are constructed
in such a way that they should be orthogonal to $|\varphi_{0}(t)\rangle$
under the modified Hamiltonian $H^{'}(t)$. Therefore, to form a complete set
of orthogonal states, the non-adiabatic state $|\varphi{}_{\pm}(t)\rangle$
must satisfy the following conditions: 
(i) $\langle|\varphi{}_{i}(t)|\varphi{}_{j}(t)\rangle=\delta_{ij}$
where $i,j=\pm,0$ (ii) $\sum_{n=0,\pm}|\varphi_{n}\rangle\langle\varphi_{n}(t)|=1$ \cite{baksic2016speeding}.
Then the chosen dressed states become
\begin{equation}
\begin{split}
|\varphi_{\pm}(t)\rangle = \frac{1}{\sqrt{2}} \Big\{ &
\big[ \sin\alpha_{2}(t) \mp \sin\alpha_{2}(t) \cos\alpha_{1}(t) \big] |00\rangle \\
&\mp \cos\alpha_{2}(t) \, |11\rangle \\
&- \big[ \cos\alpha_{1}(t) \pm i\,\sin\alpha_{2}(t) \sin\alpha_{1}(t) \big] |10\rangle
\Big\}
\label{eqn 38}
\end{split}
\end{equation}

\begin{figure*}[t]
  \centering
  \begin{minipage}[t]{0.35\textwidth}
    \centering
    \includegraphics[width=\linewidth]{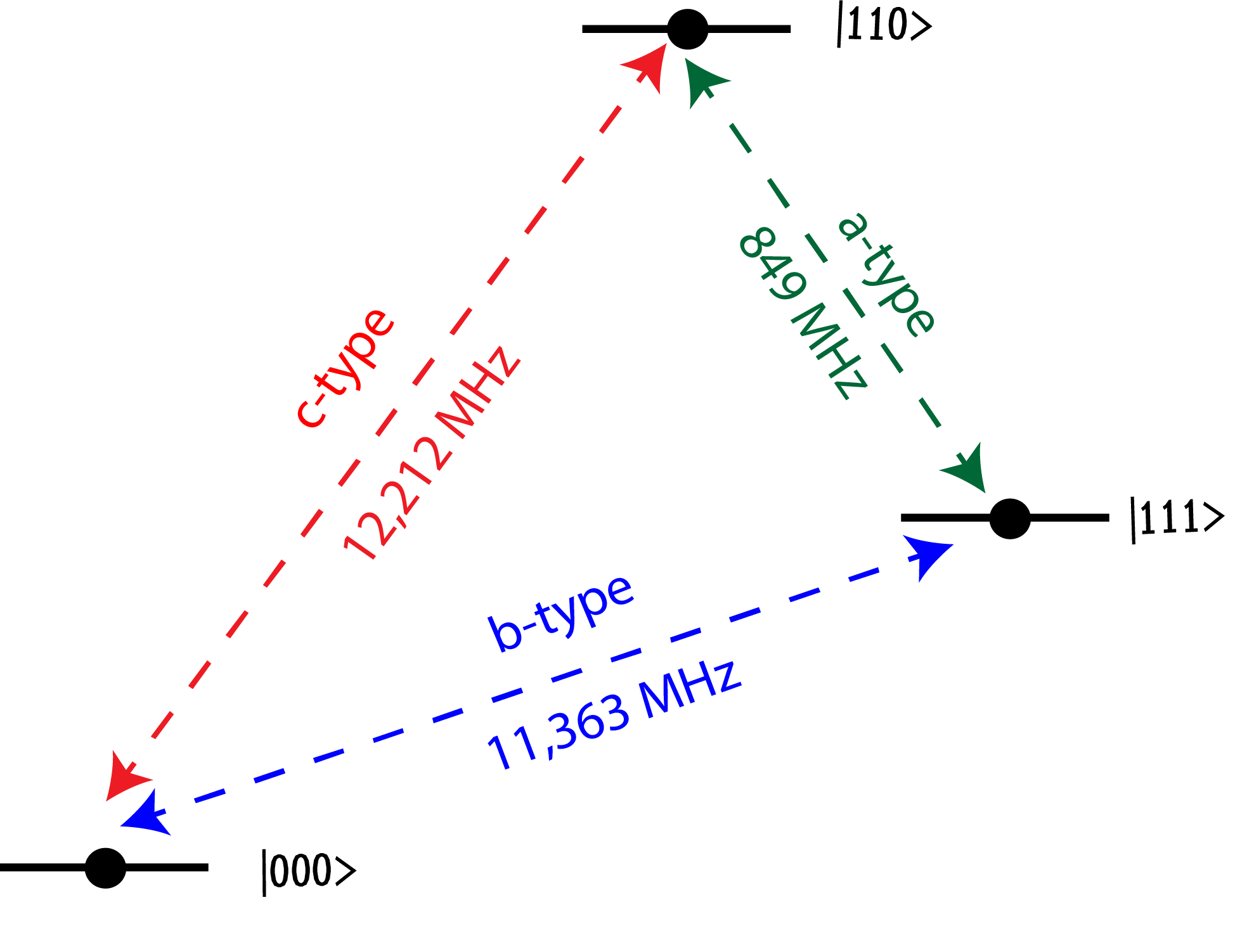}
    \caption{Allowed transition pulses $\Omega_{Q}(t)$, $\Omega_{P}(t)$, and $\Omega_{S}(t)$ in the chiral molecule 1,2-propanediol (propylene glycol, $C_{3}H_{8}O_{2}$).}
    \label{fig:06}
  \end{minipage}
  \hspace{0.15\textwidth} 
  \begin{minipage}[t]{0.45\textwidth}
    \centering
    \includegraphics[width=\linewidth]{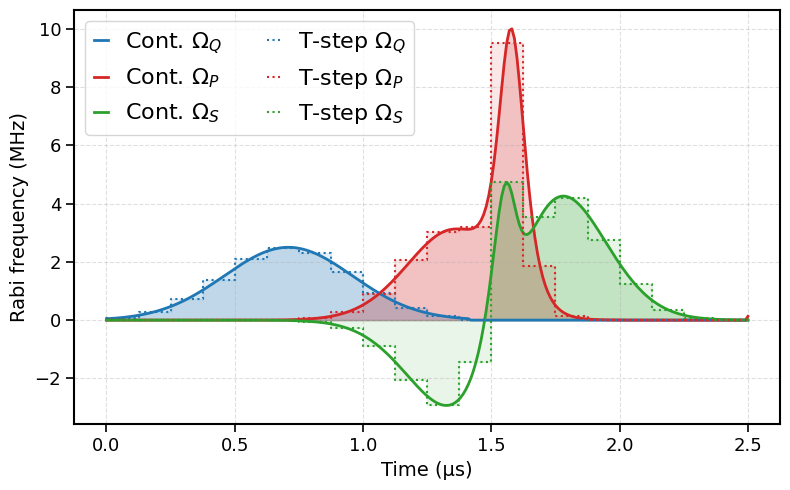}
    \caption{Continuous Gaussian envelopes after adding counteradiabatic pulses: 
    $\Omega_{Q}(t)$ (blue), 
    $\Omega_{P}(t)+\Omega_{P}'(t)$ (red), 
    $\Omega_{S}(t)+\Omega_{S}'(t)$ (green). 
    The dashed line shows the Trotterized pulses with step size $\delta t$, preserving the total pulse area.}
    \label{fig:08}
  \end{minipage}
\end{figure*}

The auxiliary controlled pulses are designed to guide $|\varphi_{0}(t)\rangle$ along the chosen path by decoupling it from the non-adiabatic states $|\varphi_{\pm}(t)\rangle$. Following the same approach we used to transform Hamiltonian $H_{PS}(t)$ into the adiabatic frame, applying $U_{1}(t)=\sum_{n=0,\pm}|\varphi_{n}\rangle\langle\varphi_{n}(t)|$
to the Hamiltonian given in Eq. (\ref{eqn 36}).

The Hamiltonian in the adiabatic frame takes the form

\begin{equation}
H'_{AD}(t)=U_{1}(t)H^{'}(t)U_{1}^{\dagger}(t)-iU_{1}(t)\dot{U}_{1}^{\dagger}(t)\label{eqn 39}
\end{equation}

{\small Both terms $\left\{ U_{1}(t)H^{'}(t)U_{1}^{\dagger}(t),iU_{1}(t)\dot{U}_{1}^{\dagger}(t)\right\} $ are
explicitly calculated in Appendix \ref{app:D}, and the final Hamiltonian becomes}{\small\par}

\begin{equation}
\begin{split}
H'_{AD}(t) = &\ \varUpsilon(t) 
\Big[ |\varphi_{+}\rangle\langle\varphi_{+}| 
      - |\varphi_{-}\rangle\langle\varphi_{-}| \Big] \\
&+ \Big[ \lambda_{+}(t) \, |\varphi_{+}(t)\rangle\langle\varphi_{0}(t)| 
      + \lambda_{-}(t) \, |\varphi_{-}(t)\rangle\langle\varphi_{0}(t)| 
      + \text{h.c.} \Big]
\label{eqn 40}
\end{split}
\end{equation}

With three parameters: 
\begin{equation*}
\begin{split}
\varUpsilon(t) 
&= \Big[ 
      \Omega_{S}'(t)\cos\alpha_{1}(t) 
      - \Omega_{P}' (t)\sin\alpha_{1}(t) \\
& \hspace{3.5em}
      + \Omega \cos 2\alpha_{1}(t) 
    \Big] 
    \cos\alpha_{2}(t)
\end{split}
\end{equation*}

\begin{align*}
\lambda_{\pm}(t) 
=\ & i \Big[ 
      \dot{\alpha}_{1}(t)\cos\alpha_{2}(t) 
      - \sin\alpha_{2}(t)\Big( 
          \Omega_{S}'(t)\cos\alpha_{1}(t) \\
& \hspace{6em}
          - \Omega_{P}' (t)\sin\alpha_{1}(t) 
          + \Omega \cos 2\alpha_{1}(t) 
      \Big) 
    \Big] \\
& \mp \Big[ 
      \Omega_{P}'(t)\cos\alpha_{1}(t) 
      - \Omega_{S}' (t)\sin\alpha_{1}(t) \\
& \hspace{6em}
      + \Omega \sin 2\alpha_{1}(t) 
      + \dot{\alpha}_{2}(t) 
    \Big]
\end{align*}

It is evident that $\lambda_{\pm}(t)$ gives information about the
coupling strength between $|\varphi_{0}\rangle\langle\varphi_{0}(t)$
and $|\varphi_{\pm}\rangle\langle\varphi_{\pm}(t)$. To guarantee
that the system only follows $|\varphi_{0}(t)\rangle$, the auxiliary
controlled pulses $\Omega_{P}^{'}(t)$ and $\Omega_{S}^{'}(t)$ are
engineered in such a way that the transition amplitude $\lambda_{\pm}(t)=0$. 

For $\lambda_{\pm}(t)=0$ and solving both equations for $\Omega_{P}^{'}(t)$ and $\Omega_{S}^{'}(t)$
gives

\begin{equation}
\begin{split}
\Omega_{P}'(t) = & -\sin\alpha_{1}(t) 
\Big[ \dot{\alpha}_{1}(t) \cot\alpha_{2}(t) + \Omega(t) \Big] \\
& - \dot{\alpha}_{2}(t) \cos\alpha_{1}(t)
\end{split}
\end{equation}

\begin{equation}
\begin{split}
\Omega_{S}'(t) = & \ \cos\alpha_{1}(t) 
\Big[ \dot{\alpha}_{1}(t) \cot\alpha_{2}(t) - \Omega(t) \Big] \\
& - \dot{\alpha}_{2}(t) \sin\alpha_{1}(t)
\end{split}
\end{equation}

Since $\hspace{0.1cm}\Omega_{S}(t)=\Omega(t)\cos\alpha_{1}(t)\hspace{0.1cm}\text{and}\hspace{0.1cm}\Omega_{P}(t)=\Omega(t)\sin\alpha_{1}(t)$

\begin{equation}
\begin{split}
\Omega_{P}(t) + \Omega_{P}'(t) = 
& - \dot{\alpha}_{1}(t) \sin\alpha_{1}(t) \cot\alpha_{2}(t) \\
& - \dot{\alpha}_{2}(t) \cos\alpha_{1}(t)
\end{split}
\label{eqn 36}
\end{equation}

and

\begin{equation}
\Omega_{S}(t)+\Omega_{S}'(t)=\dot{\alpha_{1}}(t)\cos\alpha_{1}(t)\cot\alpha_{2}(t)-\dot{\alpha_{2}}(t)\sin\alpha_{1}(t)
\label{eqn 37}
\end{equation}

By employing the modified Rabi frequency given in the Eq. (\ref{eqn 36}) and  (\ref{eqn 37}), the system
follows the three level multi-photon path $\{|\varphi_{0}(t)\rangle,|\varphi_{\pm}(t)\rangle\}$. Once the superposition {$|\Psi\rangle_{L}=\frac{1}{\sqrt{2}}(|00\rangle_{L}-|10\rangle_{L})$
and $|\Psi\rangle_{R}=\frac{1}{\sqrt{2}}(|00\rangle_{R}+|10\rangle_{R})$
is created by applying $\Omega_{Q}(t)$ pulse, the modified Rabi frequency
will drive the L and R molecules to their respective final states:

\[
|\Psi(t=t_{f})\rangle_{L}=-|10\rangle_{L})
\]
\[
|\Psi_{R}(t=t_{f})\rangle=\cos\rho^{c}|00\rangle+\sin\rho^{c}|11\rangle
\]

Where $\rho^{c}=\frac{1}{2}(\int\varUpsilon(t)dt)$ over the time
interval $t_{0}\rightarrow t_{f}$.

\subsection{Real Molecule: 1,2-propanediol}
\label{sec 02}
For a general asymmetric molecule \cite{gordy1984microwave}, the rotational energy levels are
governed by a rigid rotor Hamiltonian

\begin{equation}
\hat{H}_{rot}=A\hat{J}_{A}^{2}+B\hat{J}_{B}^{2}+C\hat{J}_{C}^{2}
\end{equation}

where $A=\frac{\hbar^{2}}{2I_{A}},\hspace{0.1em}B=\frac{\hbar^{2}}{2I_{B}},\hspace{0.1em}C=\frac{\hbar^{2}}{2I_{C}}$
and $I_{A}\neq I_{B}\neq I_{C}$

The angular momentum operators {$\hat{J}_{A},\hspace{0.1em}\hat{J}_{B},\hspace{0.1em} \text{and}\hspace{0.1em}\hat{J}_{C}$} account for the principal axis which serves as a good quantum number, but asymmetrical molecules do not conserve the projection quantum number $K$. In symmetric top molecules, $K$ is the quantum
number associated with the projection of total angular momentum $\vec{J}$
onto a molecule-fixed axis. If the molecule-fixed axis is the $ z$-axis, then $\hat{J}_{z}$ commutes with the Hamiltonian, conserving the projection quantum number $K$.

\[
\left[\hat{H},\hat{J}_{z}\right]=0\hspace{1em}\implies K=constant
\]

As for asymmetric top molecules, $I_{A}\neq I_{B}\neq I_{C}$ , none
of the components of {$\hat{J}_{A},\hspace{0.1em}\hat{J}_{B},\hspace{0.1em} \text{and}\hspace{0.1em}\hat{J}_{C}$}
commute with the Hamiltonian $\hat{H}_{rot}$

\[
\left[\hat{H}_{rot},\hat{J}_{A}\right]\neq0,\hspace{1em}\left[\hat{H}_{rot},\hat{J}_{B}\right]\neq0\hspace{1em}\left[\hat{H}_{rot},\hat{J}_{C}\right]\neq0
\]

Therefore, no projection of the total angular moment $\vec{J}$ along the molecule-fixed axis is conserved. To label the rotational energy level for an asymmetric
top we still use $|J_{K_{a}, K_{c}}\rangle$ where $K_{a}, K_{c}$ are projections of $\vec{J}$ along the principal axes $a$ and $c$.
We consider 1,2-propanediol (propylene glycol, $C_{3}H_{8}O_{2}$) as an asymmetric molecule to verify this
scheme using a quantum computer and discuss its future implications.

In the present scheme, we consider a closed-loop configuration  $|1\rangle\leftrightarrow|2\rangle\leftrightarrow|3\rangle\leftrightarrow|1\rangle$. 
The rotational energy level $|J_{K_{a}, K_{c}}\rangle$ is encoded in a two-qubit representation,  
defined by the computational basis states $\{|00\rangle,|01\rangle,|10\rangle,|11\rangle\}$. From the experimental microwave spectroscopy \cite{patterson2013enantiomer}, the rotational constants
for a conformer of 1,2-Propanediol ($C_3H_8O_2$) are: $A=8572.05MHz,\hspace{0.5em}B=3640.11MHz$
and $C=2790.97MHz$. The electric dipole moment $\vec{\mu}$ has nonzero projections along all three principal 
axes: $\mu_{a}=1.201\,D$, $\mu_{b}=1.916\,D$, and $\mu_{c}=0.365\,D$. These components determine the selection rules for $a$-, $b$-, and $c$-type transitions, respectively.
For a closed-loop system, the mapping of qubit states to rotational energy states is
\begin{equation}
\label{Eq.35}
\begin{aligned}
|00\rangle &:\quad J=0,\; K_{a}=0,\; K_{c}=0, \\
|11\rangle &:\quad J=1,\; K_{a}=1,\; K_{c}=1, \\
|10\rangle &:\quad J=1,\; K_{a}=1,\; K_{c}=0.
\end{aligned}
\end{equation}
Crucially, this mapping identifies $\ket{00}$ as the common physical ground state ($J=0$) for both enantiomers, confirming that they share an identical initialization vector at $t=0$.
 The transition frequencies \cite{lovas2009microwave}, which correspond to the energy differences between the respective states (and hence the microwave frequencies required to drive the transitions), are given by
\begin{equation}
\label{Eq 36}
\begin{aligned}
\omega_{00,11} &=\frac{E_{|11\rangle} - E_{|00\rangle}}{\hbar}  = 11363~\text{MHz}, \\[0.5em]
\omega_{00,10} &= \frac{E_{|10\rangle} - E_{|00\rangle}}{\hbar} = 12212~\text{MHz}, \\[0.5em]
\omega_{11,10} &= \frac{E_{|10\rangle} - E_{|11\rangle}}{\hbar} = \;\;849~\text{MHz}.
\end{aligned}
\end{equation}

\begin{figure*}[t!]
  \centering

  \begin{minipage}[t]{0.48\textwidth}
    \centering
    \includegraphics[width=\linewidth]{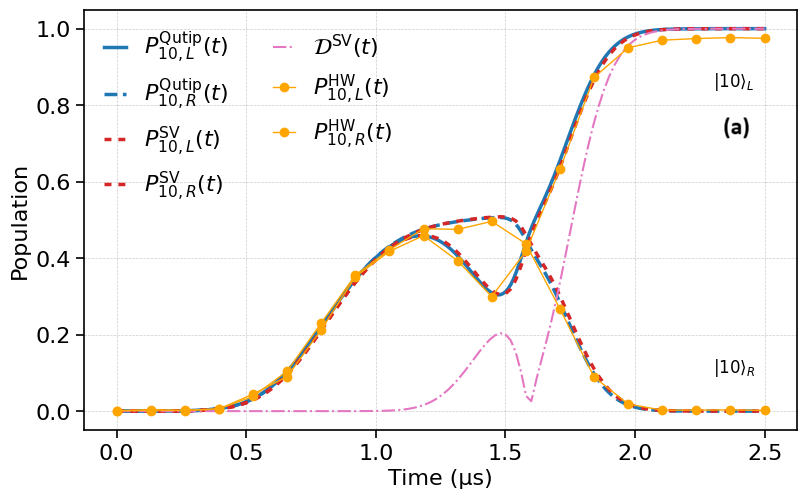}
  \end{minipage}\hfill
  \begin{minipage}[t]{0.48\textwidth}
    \centering
    \includegraphics[width=\linewidth]{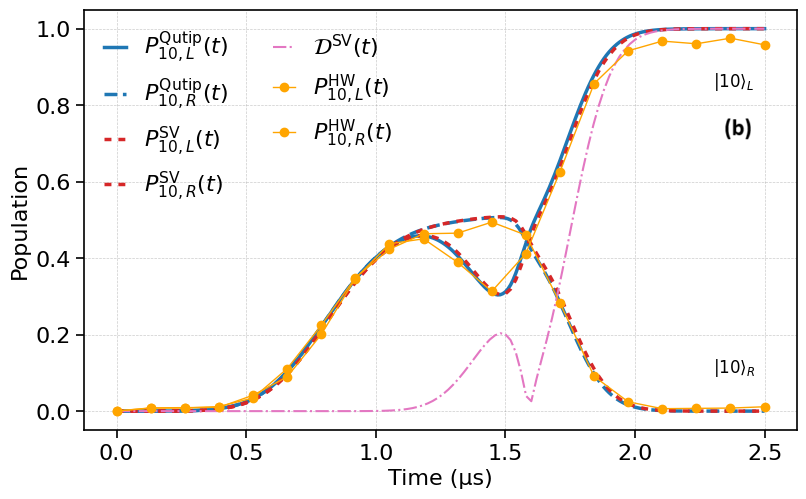}
  \end{minipage}

  \caption{Chiral discrimination of L and R enantiomers for 1,2-Propanediol ($C_3H_8O_2$), using the Shortcuts-to-Adibabcity Passage (STAP) scheme. Exact diagonalization with QuTiP (blue solid), the statevector simulator (red dashed), and quantum processor experiments (yellow solid) are shown. Panels (a) and (b) show benchmarking of the exact diagonalization results with (ibm\_kingston and ibm\_fez) quantum processors.}
  \label{fig:09}
\end{figure*}

\begin{figure*}[t!]
  \centering

  \begin{minipage}[t]{0.48\textwidth}
    \centering
    \includegraphics[width=\linewidth]{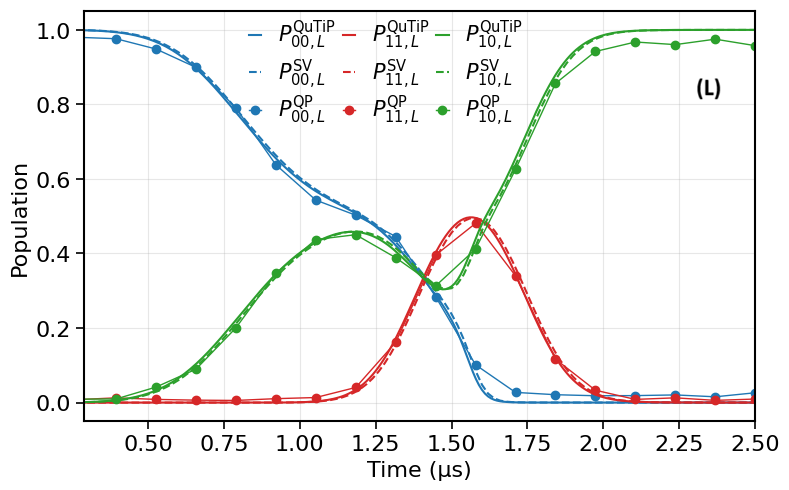}
  \end{minipage}\hfill
  \begin{minipage}[t]{0.48\textwidth}
    \centering
    \includegraphics[width=\linewidth]{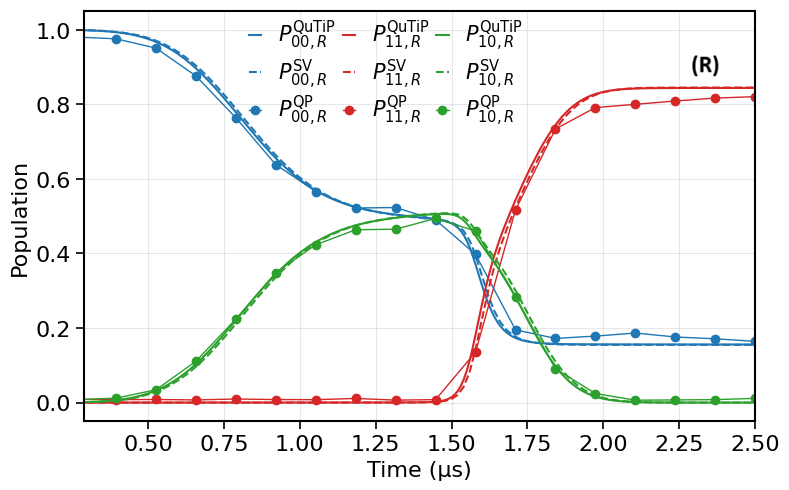}
  \end{minipage}

  \caption{Time evolution of L and R enantiomers under Shortcuts-to-Adibabcity Passage (STAP) protcol for $|00\rangle\hspace{0.1cm}\text{(blue)}$, $|11\rangle\hspace{0.1cm}\text{(red)}$, and $|10\rangle\hspace{0.1cm}\text{(green)}$ states: (i) exact diagonalization using QuTip $(P^{(\text{QuTip})})$, (ii) quantum circuit simulation performed with the statevector simulator $(P^{(\text{SV})})$, and (iii) benchmarking of the exact diagonalization results with quantum processors $(P^{(\text{QP})})$.}
  \label{fig:10}
\end{figure*}

For the 1,2-propanediol closed-loop system, the c-type transition is driven by the $\Omega_{Q}(t)$ pulse between $|00\rangle \leftrightarrow |10\rangle$ with frequency
$\omega_{00,10}=12212MHz$ . Owing to the opposite signs of this transition in the two enantiomers, it becomes the key mechanism for discrimination.
The a-type transition is driven by $\Omega_{S}(t)$
pulse between $|11\rangle\leftrightarrow|10\rangle$ with frequency
$\omega_{11,10}=849MHz$ and b-type transition is driven resonantly
by $\Omega_{P}(t)$ between $|00\rangle\leftrightarrow|11\rangle$
with frequency $\omega_{00,11}=11363MHz$. 
The Rabi frequency $\Omega_{j}(t)$ for each transition $(j=P,Q,S)$
is determined by the interaction between the transition dipole moment $\mu_{j}$ and the applied electric field $\epsilon_{j}$

\begin{equation}
\Omega_{j}(t)=\frac{\vec{\mu_{j}}.\vec{\epsilon_{j}}}{\hbar}
\end{equation}
This relationship constitutes the precise mechanism for encoding the physical problem instance into the quantum processor. The process operates in two steps: (1) \textbf{Allocation:} The specific spectroscopic transitions (a-, b-, and c-type) defined in Eq. \ref{Eq 36} are allocated to the specific qubit operators derived in Sec. \ref{sec III} (e.g., the c-type transition maps to the Q-pulse gate acting on the $\ket{00}\leftrightarrow\ket{10}$ subspace). (2) \textbf{Encoding:} The molecule's unique dipole moments ($\vec{\mu}_j$) determine the instantaneous Rabi frequencies $\Omega_j(t)$, which are directly encoded as the rotation angles $\theta_i = \Omega_j(t_i)\delta t$ in the Trotterized gate sequence. Thus, the hardware execution is strictly constrained by the physical parameters of the 1,2-propanediol instance.}
The field strengths of $0$–$2 \,\text{V/cm}$ produce the following Rabi frequencies: $\Omega_{P}(t) \approx 10$–$12 \,\text{MHz}$ for the $|00\rangle \leftrightarrow |11\rangle$ transition, $\Omega_{S}(t) \approx 5$–$8 \,\text{MHz}$ for the $|11\rangle \leftrightarrow |10\rangle$ transition, and $\Omega_{Q}(t) \approx 2.5$–$3 \,\text{MHz}$ for the $|00\rangle \leftrightarrow |10\rangle$ transition, as shown in Fig.~\ref{fig:06}. These frequencies ensure the phase stability
which is critical for enantiomer discrimination and coherent population
transfer without higher power distortions. Most importantly, the field
strength of $0\sim2V/cm$ is experimentally accessible and yields
$MHz$ scale $\Omega_{j}(t)$, which satisfies the Rotating-Wave approximation
$(\Omega_{j}(t)\ll\omega_{j})$.

Similarly, for STAP, we used Trotter Decomposition {
to divide the continuous Gaussian pulses into discrete rectangular shapes
as shown in Fig. \ref{fig:08}. The $\Omega_{Q}(t)$ pulse begins at $t=0$ and peaks at $t\approx 1.5\,\mu s$, driving L and R enantiomers from the initial state $|00\rangle$ to $|\Psi\rangle_{L}=\tfrac{1}{\sqrt{2}}(|00\rangle_{L}-|10\rangle_{L})$ and $|\Psi\rangle_{R}=\tfrac{1}{\sqrt{2}}(|00\rangle_{R}+|10\rangle_{R})$, respectively.
 Two parameters, $\alpha_{1}(t)$ and $\alpha_{2}(t)$, mainly control the time and shape of the pulses in STAP. In particular, $\alpha_{1}(t)$ controls the superposition between states $|00\rangle$ and $|10\rangle$, analogous to the STIRAP mixing angle. Meanwhile, $\alpha_{2}(t)$ regulates the intermediate-state population $|11\rangle$,
functioning as a counter-adiabatic controller. In STIRAP, the adiabatic condition $|\frac{\dot{\theta}}{\Omega(t)}|\ll1$
requires slowly varying pulses (long interaction times $T$). 
However, in the STAP protocol, the pulse shapes adapt dynamically 
to the time derivatives of $\alpha_{1}(t)$ and $\alpha_{2}(t)$.

\begin{figure*}[t!]
  \centering

  \begin{minipage}[t]{0.48\textwidth}
    \centering
    \includegraphics[width=\linewidth]{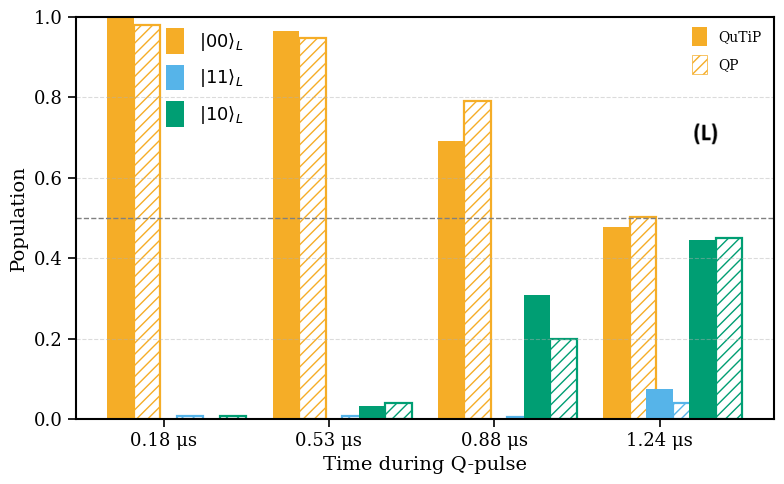}
  \end{minipage}\hfill
  \begin{minipage}[t]{0.48\textwidth}
    \centering
    \includegraphics[width=\linewidth]{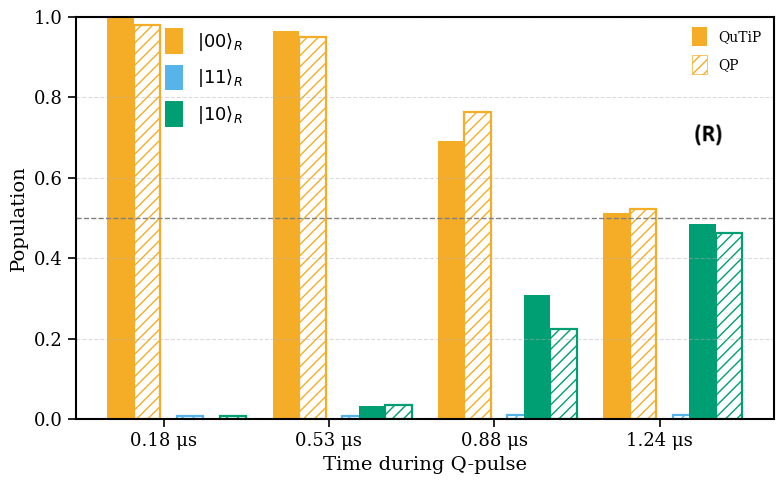}
  \end{minipage}

  \caption{State populations $|00\rangle$ (orange), $|11\rangle$ (blue), and $|10\rangle$ (green) at the midpoint ($t=0.61\,\mu s$) and end ($t=2.53\,\mu s$) of the $\Omega_{Q}(t)$ pulse, under Shortcuts to Adiabatic Passage (STAP) scheme. Bars with hatched fill represent results from the quantum processor, while solid bars show exact diagonalization using QuTiP, for both enantiomers.}
  \label{fig:11}
\end{figure*}

For instance, near the intermediate state 
$(\alpha_{1}\!\approx\!\tfrac{\pi}{4})$ of the STIRAP pulse sequence, 
$\alpha_{1}(t)$ varies most rapidly, making the system particularly 
susceptible to nonadiabatic transitions. At this point, the scheme applies a counter-adiabatic correction
 when $\dot{\alpha_{1}}(t)$ is large. Consequently, rapid variations in $\alpha_{1}(t)$ are compensated by an 
appropriately tuned counter-adiabatic correction $\alpha_2(t)$, 
which removes the adiabatic constraint and enables short pulse durations 
($T\!\approx\!10\,\mu s$ to $T\!=\!2.5\,\mu s$), as shown in Fig.~\ref{fig:09}. Following the boundary conditions at time
$t_{i}$ and $t_{f}$, we require $\alpha_{2}(t_{i})=\alpha_{2}(t_{f})=0$,
and the engineered eigenstates must match the original adiabatic
states. We can ensure these boundary conditions with a Gaussian function: $\alpha_{2}(t)=\alpha_{m}(t)e^{-[(t-t_{i})-(t_{f}-t_{i})/2]^{2}/T_{\alpha_{2}}^{2}}$ where
$T_{\alpha_{2}}=(t_{f}-t_{i})/6$.

In Fig.~\ref{fig:09}, we show the time evolution of the target state $|10\rangle$
under the Shortcut-to-Adiabatic Passage (STAP) scheme. 
The blue solid line corresponds to the QuTiP simulation, representing the ideal 
population transfer. For the L enantiomer, the population monotonically increases 
to $P_{L\rightarrow|10\rangle}=|\langle\Psi_{L}|10\rangle|^{2}\approx 1.0$, 
demonstrating perfect transfer under STAP-accelerated adiabatic conditions. The red dashed line shows the statevector simulator results, which deviate slightly
from the QuTiP curve due to finite Trotterization and pulse–discretization
approximations. Nevertheless, the agreement validates the protocol’s robustness
to Trotter error.
The yellow dotted line represents the experimental benchmark obtained from two
IBM quantum processors. As illustrated in Fig.~\ref{fig:09}(a), ibm\_kingston processor achieves
greater accuracy than ibm\_fez Fig.~\ref{fig:09}(b), primarily
because of its longer coherence times. Despite hardware limitations, the STAP protocol maintains a
discrimination fidelity of
$D(t) \approx 0.7\text{--}0.8$, underscoring its robustness and
potential applicability on current quantum hardware.

As shown in Fig.~\ref{fig:04}, STAP achieves enantio-discrimination 
within $2.5\,\mu s$, a five-fold speedup over STIRAP ($10\,\mu s$), 
by employing analytically designed counteradiabatic pulses 
$\bigl(\Omega_{P}'(t), \Omega_{S}'(t)\bigr)$ that guarantee adiabatic-like evolution. Such improvements in speed and precision are particularly valuable for the quantum simulation of large degenerate chiral molecules, where both temporal efficiency and high fidelity are essential. By significantly reducing the 
overall process time, the STAP scheme allows the quantum processor to execute simulations more accurately, enabling up to five times more  Trotter steps within the same physical coherence window.

Fig. \ref{fig:10}(L) and Fig. \ref{fig:10}(R) present the time evolution of 
$|00\rangle$, $|11\rangle$, and $|10\rangle$ states for L and R enantiomers, 
demonstrating how the counter-adiabatic fields 
$(\Omega_{P}'(t)$ and $\Omega_{S}'(t))$ guide the population transfer 
from the initial state. The green curve ($|10\rangle$) highlights the successful transfer of the L and R enantiomers 
to the target state. In contrast, the red ($|00\rangle$) and blue ($|11\rangle$) curves 
trace the depletion of the initial populations, demonstrating how the STAP protocol 
redistributes population across the three levels.
 From Fig. \ref{fig:10}, it is clear that, when a $\Omega_{Q}(t)$ pulse is applied initially to create the superposition, the  $|00\rangle$ and $|10\rangle$
achieve a higher population in their target states for both L and R, which
shows that the STAP actively suppressed nonadiabatic transitions by
shortening the time process. In Fig.~\ref{fig:11}, the $|00\rangle$ and $|10\rangle$ states are shown as 
orange and green bars, while 
the blue bars correspond to $|11\rangle$ state. Bars with hatched fill represent results from the quantum processor, while solid bars show exact diagonalization using QuTiP, for both enantiomers. Remarkably, STAP produces nearly 
ideal superpositions, 
$|\Psi_{L}\rangle=\tfrac{1}{\sqrt{2}}(|00\rangle_{L}-|10\rangle_{L})$ and 
$|\Psi_{R}\rangle=\tfrac{1}{\sqrt{2}}(|00\rangle_{R}+|10\rangle_{R})$, 
since the counter-adiabatic fields 
$\bigl(\Omega_{P}'(t),\Omega_{S}'(t)\bigr)$ act early to suppress unwanted 
couplings and stabilize the dressed-state pathway.
At $t=2.53~\mu\text{s}$, once the $\Omega_{Q}(t)$ pulse is fully applied, 
STIRAP leaves the system with populations of $0.82$ in $\ket{00}$ and $0.18$ 
in $\ket{10}$ for both enantiomers. By contrast, STAP achieves the desired 
superposition much faster, yielding nearly equal probabilities of $0.52$ and 
$0.48$ at $t=1.24~\mu\text{s}$.

\section{Scalability and Error Analysis}

Although the present demonstration uses a three-level molecular manifold,
the approach can be naturally extended to larger systems. For two
qubits, the joint Hilbert space has dimension $2^{2}=4$ with the following basis set $\{|00\rangle,|01\rangle,|10\rangle,|11\rangle\}$. In
general, n qubits have $2^{n}$ basis states. So, if a molecule has L energy
levels, we need enough qubits such that $2^{2}\geq L.$ Taking the log base 2 gives: $n\geq log_{2}L$.
which shows that the total qubit number increases only logarithmically with the number
of molecular states. In our approach, each dipole-allowed transition
is driven by a shaped pulse of the same analytical form as the Q--P--S
set used for the three-level demonstration. Because molecular selection
rules connect only neighboring states $\Delta j=\pm1$, the interaction
Hamiltonian remains sparse, so the number of required pulses grows
linearly with the number of levels. Each new level can thus be incorporated
by reusing the same STIRAP/STAP pulse module without re-optimizing
the global dynamics.The same sequence of driven transitions (Q, P,
S pulses) generalizes to additional couplings between neighboring
states determined by dipole-selection rules. Because these couplings
form a sparse network, the total number of gates increases approximately
linearly with the number of allowed transitions rather than with $L^{2}.$
This combination of logarithmic state representation and linear growth
in operational modules is far more efficient than the quadratic scaling
of classical simulations, establishing a compelling pathway for simulating
complex chiral molecules on future quantum processors.

When amplitude noise is introduced, its impact on population transfer can be analyzed via the instantaneous mixing angle
\begin{equation}
\alpha_{1} = \tan^{-1}\!\left(\frac{\Omega_{P}}{\Omega_{S}}\right),
\end{equation}
such that the dark state under amplitude fluctuations becomes
\begin{equation}
\ket{\tilde{\gamma}_{0}} =
\cos(\alpha_{1} + \delta\alpha_{1})\ket{00}
- \sin(\alpha_{1} + \delta\alpha_{1})\ket{10}.
\end{equation}
For small fractional errors in the Rabi frequencies,
\begin{equation}
\Omega_{P} \rightarrow \Omega_{P}(1+\delta_{P}), 
\quad 
\Omega_{S} \rightarrow \Omega_{S}(1+\delta_{S}),
\end{equation}
the mixing angle is perturbed by
\begin{equation}
\delta\alpha_{1} =
\frac{\Omega_{P}\Omega_{S}}{\Omega_{P}^{2} + \Omega_{S}^{2}}
(\delta_{P} - \delta_{S}).
\end{equation}
Hence, only the relative amplitude mismatch between the two pulses contributes to the deviation.
This perturbation effectively tilts the dark state and induces a leakage error,
\begin{equation}
1 - F = \left|\braket{00|\tilde{\psi}_{0}(T)}\right|^{2} 
= \left|\delta\alpha_{1}(T)\right|^{2},
\end{equation}
which scales quadratically with $(\delta_{P} - \delta_{S})$.
The detail
derivation of this probability is given in the Appendix \ref{app:E}.\\ 
In conventional STIRAP, the transfer process is first-order insensitive but operates in the slow adiabatic limit, where time-dependent fluctuations accumulate throughout the long pulse sequence. In typical shortcut-to-adiabaticity (STA) implementations, the fast counter-diabatic channel must be precisely matched to the main drive; any amplitude mismatch reintroduces first-order errors across the entire evolution, resulting in a large prefactor within the same quadratic scaling law. In contrast, our Q--P--S (STAP-digital) protocol realizes each coupling as a short, locally calibrated rotation. By embedding composite-pulse corrections within each block\cite{gevorgyan2021ultrahigh,shi2020robust,shi2021robust,kabytayev2014robustness}, the linear amplitude error cancels, and the leading infidelity scales as $1 - F \propto (\delta_{P} - \delta_{S})^{4}.$ The brief, modular segments also strongly suppress the dynamical accumulation of noise. Mathematically, this elevates the error order and reduces the effective prefactor in the fidelity loss, demonstrating that our scheme remains substantially more robust to amplitude fluctuations than both adiabatic STIRAP and standard STA approaches.}

The use of Trotter dynamics in our protocol discretizes the continuous
adiabatic or shortcut evolution into a finite sequence of gate-level
segments (Q--P--S blocks). This digital decomposition provides
several advantages when realistic parameter fluctuations are considered.
First, each segment can be locally calibrated, allowing compensation
of amplitude or phase drifts on a block-by-block basis, whereas in
a continuous process small errors accumulate coherently over the full
evolution time. Second, the discrete structure permits the insertion
of composite or echo sub-pulses that cancel first-order amplitude
and detuning errors without altering the target evolution, thereby
improving robustness to slow control noise. In addition, Trotterization
maps naturally onto digital quantum hardware, where native gate sets
are inherently discrete and can be optimized individually. The primary
drawback, however, is the introduction of Trotter (digitization) error:
finite segmentation approximates the continuous Hamiltonian only up
to $O((\Delta t)^{2})$corrections, which can slightly distort the
ideal adiabatic trajectory if the number of steps is too small. Moreover,
sharp transitions between segments can increase sensitivity to high-frequency
control noise or bandwidth limitations. Overall, the Trotterized approach
exchanges a small, controllable discretization error for significant
gains in calibration flexibility, noise compensation, and hardware
compatibility---yielding higher practical fidelity under realistic
parameter fluctuations compared with long, continuous driving schemes.

\section{Conclusion}

We have demonstrated a chiral discrimination scheme for the real molecule 1,2-propanediol using  Stimulated Raman Adiabatic Passage (STIRAP), mapped onto a two-qubit quantum circuit, and experimentally benchmarked on IBM quantum processors. Under adiabatic conditions, the STIRAP protocol is inherently slow, allowing unwanted transitions that cause population leakage into intermediate states. As a result, it fails to efficiently transfer the population into the target states, making it unsuitable for experimental chiral discrimination. In contrast, implementing the Shortcuts-to-Adiabatic Passage (STAP) protocols markedly reduces evolution time and achieves higher-fidelity population transfer, resulting in clear separation of L and R enantiomers. These results establish a foundation for employing gate-based quantum computing to study and control molecular chirality, paving the way for investigations of more complex systems with degenerate energy states and developing quantum algorithms to probe enantiomer-specific interactions \cite{brif2010control}. Beyond demonstrating chiral discrimination via STAP on gate-based platforms, a promising avenue is using quantum circuits to actively control molecular reaction pathways and unravel the complexity of chemical quantum dynamics \cite{judson1992teaching,brumer1986control,koch2019quantum}.

The goal of simulating chiral discrimination on a gate-based quantum processor is not to test  computational speed, but to examine whether the enantiomer-selective physics of the three-level system survives when continuous control is replaced by a finite sequence of native gates under realistic noise. Our circuit mapping preserves the dark state structure and relative phases while discretizing the pulses, and hardware experiments show that even a modest number of Trotter slices can generate the expected Left/Right  contrast, demonstrating that the discriminator is inherently robust in a digital setting. Running the protocol on an actual quantum processor, rather than a simulator, reveals how the signal withstands real noise and identifies the imperfections such as amplitude miscalibration or dephasing that most strongly affect performance. This provides a realistic estimate of the gates, qubits, and measurements required for success. Beyond confirming the theory, these results show that our method for chiral discrimination remains effective under real-world noise and operates efficiently within the resource limits of current quantum hardware, thereby bridging a theoretical concept to a scalable quantum application.

\section{ACKNOWLEDGMENTS}
The authors would like to acknowledge the financial support from the Quantum Science Center, a National Quantum Information Science Research Center of the U.S. Department of Energy (DOE), operated at Oak Ridge National Laboratory (ORNL).

\clearpage
\bibliographystyle{apsrev4-2}
\bibliography{bibliography}   

\clearpage
\appendix
\onecolumngrid

\setcounter{equation}{0}
\makeatletter
\renewcommand{\theequation}{A\arabic{equation}}
\makeatother

\refstepcounter{section}            
\section*{APPENDIX \Alph{section}: Derivation of the Interacting Hamiltonian}
\phantomsection                      
\label{app:A}

Here, we present the derivation of the  Hamiltonian in the interaction picture, $H_{int}(t)=e^{iH_{0}t}H_{1}e^{-iH_{0}t} $, by substituting the explicit forms of $H_{0} $ and $H_{1}$. For the transition $|1\rangle\Leftrightarrow|2\rangle$
with energy splitting $\omega_{12}$, the dipole interaction term written as

\begin{equation}
H_{int}(t) = e^{iH_{0}t} 
\Bigg[ \frac{\Omega_{P}(t)}{2} 
\Big( e^{i(\omega_{P}t+\phi_{P})} + e^{-i(\omega_{P}t+\phi_{P})} \Big) 
\left( |1\rangle\langle 2| + |2\rangle\langle 1| \right)
\Bigg] e^{-iH_{0}t}
\label{eqn:07}
\end{equation}

\begin{equation*}
H_{int}(t) = \frac{\Omega_{P}(t)}{2} \Big[
e^{i(\omega_{P} - \omega_{12})t} e^{i\phi_{P}} \, |1\rangle\langle 2|
+ e^{-i(\omega_{P} - \omega_{12})t} e^{-i\phi_{P}} \, |2\rangle\langle 1|
\Big]
\end{equation*}

Assume that the applied field is nearly resonant $(\omega_{P}\approx\omega_{12})$

\begin{equation}
H_{int}(t)=\frac{\Omega_{P}(t)}{2}\left[e^{i\phi_{P}}|1\rangle\langle2|+e^{-i\phi_{P}}|2\rangle\langle1|\right]\label{eqn 08}
\end{equation}

In the case of a $|2\rangle\Leftrightarrow|3\rangle$ transition characterized
by an energy separation $\omega_{12}$, the corresponding dipole interaction
term becomes

\begin{equation*}
H_{int}(t) = e^{iH_{0}t} \Bigg[
\frac{\Omega_{S}(t)}{2} \Big(
e^{i(\omega_{S}t+\phi_{S})} + e^{-i(\omega_{S}t+\phi_{S})}
\Big)
\left( |2\rangle\langle 3| + |3\rangle\langle 2| \right)
\Bigg] e^{-iH_{0}t}
\label{eqn:09}
\end{equation*}

\begin{equation}
H_{int}(t) = \frac{\Omega_{S}(t)}{2} \Big[
e^{i\big(\omega_{S}t + (\omega_{12} - \omega_{13})\big)} e^{i\phi_{S}} \, |2\rangle\langle 3|
+ e^{-i\big(\omega_{S}t + (\omega_{12} - \omega_{13})\big)} \, |3\rangle\langle 2|
\Big]
\label{eqn:XX} 
\end{equation}

Since $\omega_{12}-\omega_{13}\approx\omega_{23}\approx\omega_{S}$, the Hamiltonian $H_{int}(t)$ becomes

\begin{equation}
H_{int}(t)=\frac{\Omega_{S}(t)}{2}\left[e^{i\phi_{S}}|2\rangle\langle3|+e^{-i\phi_{S}}|3\rangle\langle2|\right]\label{eqn 10}
\end{equation}

Similarly for the transition $|1\rangle\Leftrightarrow|3\rangle$
with energy splitting $\omega_{13}$, the dipole interaction term
takes the form

\begin{equation}
H_{int}(t) = e^{iH_{0}t} \Bigg[
\frac{\Omega_{Q}(t)}{2} \Big(
e^{i(\omega_{Q}t+\phi_{Q})} + e^{-i(\omega_{Q}t+\phi_{Q})}
\Big)
\left( |1\rangle\langle 3| + |3\rangle\langle 1| \right)
\Bigg] e^{-iH_{0}t}
\label{eqn:11}
\end{equation}

\begin{equation*}
H_{int}(t) = \pm \frac{\Omega_{Q}(t)}{2} \Big[
e^{i\big(\omega_{Q}t - \omega_{13}\big)} e^{i\phi_{Q}} \, |1\rangle\langle 3|
+ e^{-i\big(\omega_{Q}t - \omega_{13}\big)} e^{-i\phi_{Q}} \, |3\rangle\langle 1|
\Big]
\label{eqn:XX} 
\end{equation*}

Consider the applied field is nearly resonant $(\omega_{Q}\approx\omega_{13})$

\begin{equation}
H_{int}(t)=\pm\frac{\Omega_{Q}(t)}{2}\left[e^{i\phi_{Q}}|1\rangle\langle3|+e^{-i\phi_{Q}}|3\rangle\langle1|\right]\label{eqn 12}
\end{equation}

Finally, the Hamiltonian matrix in the basis set $(|1\rangle, |2\rangle,|3\rangle)$ becomes

\begin{equation}
H_{int}(t)=\frac{1}{2}\left(\begin{array}{ccc}
0 & \Omega_{P}e^{i\phi_{P}} & \pm\Omega_{Q}e^{i\phi_{Q}}\\
\Omega_{P}e^{-i\phi_{P}} & 0 & \Omega_{S}e^{i\phi_{S}}\\
\pm\Omega_{Q}e^{-i\phi_{Q}} & \Omega_{S}e^{-i\phi_{S}} & 0
\end{array}\right)\label{eqn 13}
\end{equation}

\setcounter{equation}{0}
\setcounter{figure}{0}
\makeatletter
\renewcommand{\theequation}{B\arabic{equation}}
\makeatother
\renewcommand{\thefigure}{B\arabic{figure}} 
\refstepcounter{section}  
\section*{APPENDIX \Alph{section}: Quantum Gates}
\phantomsection           
\label{app:B}

\setcounter{subsection}{0}
\renewcommand{\thesubsection}{\Roman{subsection}}

\subsection{\texorpdfstring{$Q$-pulse}{Q-pulse}}
The Hamiltonian for $Q-$pulse only is

\begin{equation}
\begin{split}
H_{int}(t)=\pm\frac{\Omega_{Q}(t)}{2}\left[e^{i\phi_{Q}}|00\rangle\langle10|+e^{-i\phi_{Q}}|10\rangle\langle00|\right]
\label{eqn 14a}
\end{split}
\end{equation} 

We use the tensor product rule: $\hspace{0.1cm}|ab\rangle\langle cd| \;=\; (\,|a\rangle\langle c|\,)_{q_{0}} \;\otimes\; (\,|b\rangle\langle d|\,)_{q_{1}} .$

\vspace{0.5cm}
$\implies$

\[
\begin{aligned}
|00\rangle\langle 10| = (\,|0\rangle\langle 1|\,)_{q_{0}} \otimes (\,|0\rangle\langle 0|\,)_{q_{1}} \hspace{0.1cm}\\ 
|10\rangle\langle 00| = (\,|1\rangle\langle 0|\,)_{q_{0}} \otimes (\,|0\rangle\langle 0|\,)_{q_{1}} 
\end{aligned}
\]

Therefore
\begin{equation}
H_{Q}(t_{i}) \;=\; \pm \frac{\Omega_{Q}(t_{i})}{2} \Big[
e^{i\phi_{Q}} \,(|0\rangle\langle 1|)_{q_{0}} \otimes (|0\rangle\langle 0|)_{q_{1}}
\;+\;
e^{-i\phi_{Q}} \,(|1\rangle\langle 0|)_{q_{0}} \otimes (|0\rangle\langle 0|)_{q_{1}}
\Big] .
\end{equation}

\[
e^{i\phi_Q}\ket{0}\bra{1} + e^{-i\phi_Q}\ket{1}\bra{0}
= \cos\phi_Q (\ket{0}\bra{1} + \ket{1}\bra{0})
+ i\sin\phi_Q (\ket{1}\bra{0} - \ket{0}\bra{1})
\]

As
\[
X=\ket{0}\!\bra{1}+\ket{1}\!\bra{0}, \qquad
Y=i\bigl(\ket{1}\!\bra{0}-\ket{0}\!\bra{1}\bigr)
\]

\[
\Rightarrow \quad
e^{i\phi_Q}\ket{0}\bra{1} + e^{-i\phi_Q}\ket{1}\bra{0}
= \cos\phi_Q X + \sin\phi_Q Y
\]

and

\[
H_Q(t_i)
= \pm \frac{\Omega_Q(t_i)}{2}
\left[
\cos\phi_Q X_{q_0} + \sin\phi_Q Y_{q_0}
\right]
\otimes (\ket{0}\bra{0})_{q_1}
\]

For one Trotter slice:

\[
U_Q(t_i) = \exp\!\left[-i \frac{\theta_i}{2}
\left\{
\cos\phi_Q X_{q_0} + \sin\phi_Q Y_{q_0}
\right\}
\otimes (\ket{0}\bra{0})_{q_1}
\right],
\quad
\theta_i = \Omega_Q(t_i)\,\delta t
\]

Let $\theta_i = \Omega_Q(t_i)\,\delta t$ and define the single-qubit generator on $q_0$:
\[
A_{\phi} \equiv \cos\phi_Q\, X_{q_0} + \sin\phi_Q\, Y_{q_0}
\]

Then the evolution operator is
\[
U_Q(t_i) = \exp\!\left[-i \frac{\theta_i}{2}\, A_{\phi} \otimes \ket{0}\!\bra{0}_{q_1}\right]
\]

As we know, the projector spectral identity is:
$\ket{1}\!\bra{1}_{q_1} + \ket{0}\!\bra{0}_{q_1} = I$.

\[
\exp\!\left[-\,i \frac{\theta_i}{2} A \otimes \ket{0}\!\bra{0}\right]
= \ket{0}\!\bra{0} \otimes e^{-\,i \frac{\theta_i}{2} A}
+ \ket{1}\!\bra{1} \otimes I
\]

Hence
\[
U_Q(t_i)
= \ket{0}\!\bra{0}_{q_1} \otimes e^{-\,i \frac{\theta_i}{2} A_{\phi}}
+ \ket{1}\!\bra{1}_{q_1} \otimes I_{q_0}
\]

\text{This makes the condition explicit:}
\begin{itemize}
  \item if $q_1 = 0$: apply $e^{-\,i \frac{\theta_i}{2} A_{\phi}}$ to $q_0$;
  \item if $q_1 = 1$: do nothing.
\end{itemize}

The pulse phase $\phi_{Q}$ controls the rotation axis; for $\phi_{Q} = \frac{\pi}{2}$, the $Q$-pulse implements a $Y$-axis rotation.

\[
U_Q(t_i)
= \ket{0}\!\bra{0}_{q_1} \otimes R_{y}(\theta_i)
+ \ket{1}\!\bra{1}_{q_1} \otimes I_{q_0},
\quad \text{where } \theta_i = \Omega_Q(t_i)\,\delta t.
\]

Quantum hardware uses controls on $\bra{1}$, not $\bra{0}$. We use the identity

\[
\ket{0}\bra{0} = X\ket{1}\bra{1}X,
\]
we can write

\[
U_Q(t_i)
= (I \otimes X)
\left[
\ket{1}\!\bra{1}_{q_1} \otimes  R_y(\theta_i)\
+ \ket{0}\!\bra{0}_{q_1} \otimes I
\right]
(I \otimes X).
\]

Controlled-$R_{y}(\theta)$ (control $q_1$, target $q_0$) uses two CNOTs:

\[
\text{Controlled-}R_{y}(\theta)
= \mathrm{CNOT}_{q_1 \rightarrow q_0}\,
R_{y}(\theta)_{q_0}\,
\mathrm{CNOT}_{q_1 \rightarrow q_0}.
\]

Since  control is on $\ket{0}$, wrap with $X$ on $q_1$:

\[
U_Q(t_i)
= (I \otimes X)\,
\mathrm{CNOT}\,
R_{y}(\theta_i)\,
\mathrm{CNOT}\,
(I \otimes X).
\]

\begin{figure}[h!]
    \centering
    \fbox{%
        \begin{minipage}{0.85\textwidth}  
            \centering
            \includegraphics[width=0.55\textwidth]{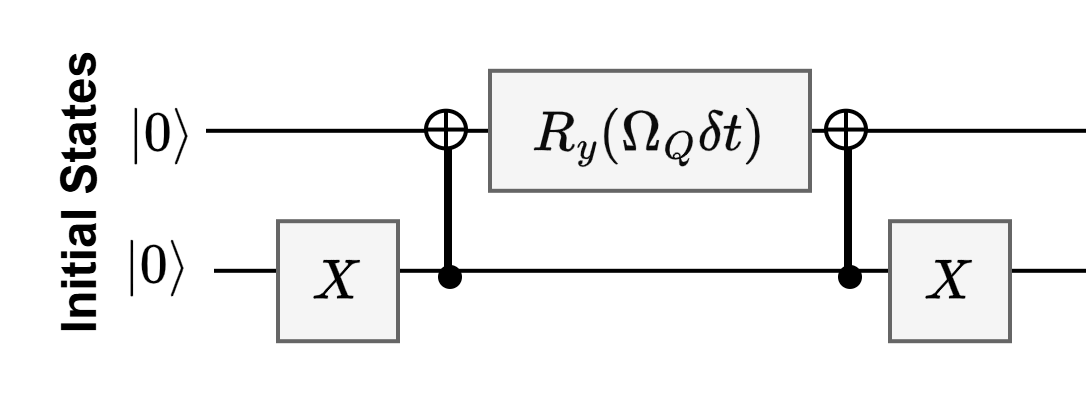} 
            \captionsetup{justification=centering}
            \caption{
            Quantum-circuit realization of the $Q$-pulse.
            The operation implements a conditional single-qubit rotation
            $R_y(\Omega_Q \delta t)$ on the upper qubit, using $X$ wrappers
            to convert the control condition from $\ket{0}$ to $\ket{1}$.
            }
            \label{fig:appB_Qpulse}
        \end{minipage}%
    }
\end{figure}

However, the controlled rotation can be implemented using a $(CX-Ry-CX)$ sequence, as illustrated in Fig. \ref{fig:appB_Qpulse}. \\

\subsection{$P-$ and $S-$pulse gate}

For $P-$pulse the Hamiltonian $H_{P}(t)$ is

\begin{equation}
H_{P}(t)=\frac{\Omega_{p}(t)}{2}\left(|00\rangle\langle11|+|11\rangle\langle00|\right)\label{eqn 19}
\end{equation}

\begin{equation}
    H_P(t)=\frac{\Omega_P(t)}{2}\left[
\bigl(\ket{0}\!\bra{1}\bigr)_{\,q_0}\otimes\bigl(\ket{0}\!\bra{1}\bigr)_{\,q_1}
+
\bigl(\ket{1}\!\bra{0}\bigr)_{\,q_0}\otimes\bigl(\ket{1}\!\bra{0}\bigr)_{\,q_1}
\right]
\end{equation}

\begin{figure}[t!]
    \centering
    \fbox{%
        \begin{minipage}{0.9\textwidth}
            \centering
            \includegraphics[width=0.85\textwidth]{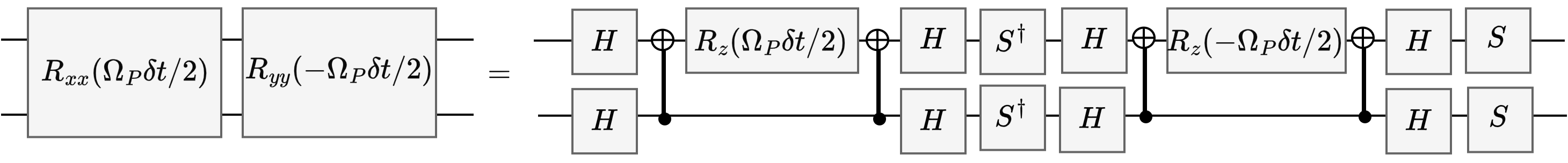} 
            \captionsetup{justification=centering}
            \caption{
            Quantum-circuit realization of the $P$-pulse.
            The operation decomposes the composite evolution 
            $R_{XX}(\Omega_P \delta t / 2) , \hspace{0.2cm} R_{YY}(-\Omega_P \delta t / 2)$ 
            into native single- and two-qubit gates.
            Hadamard ($H$) and phase ($S, S^{\dagger}$) transformations 
            convert the $R_{XX}$ and $R_{YY}$ interactions into controlled 
            $R_z$ rotations, producing an experimentally implementable 
            gate sequence equivalent to the ideal two-qubit operation.
            }
            \label{fig:appB_Ppulse}
        \end{minipage}%
    }
\end{figure}

\begin{figure}[t!]
    \centering
    \fbox{%
        \begin{minipage}{0.9\textwidth}
            \centering
            \includegraphics[width=0.85\textwidth]{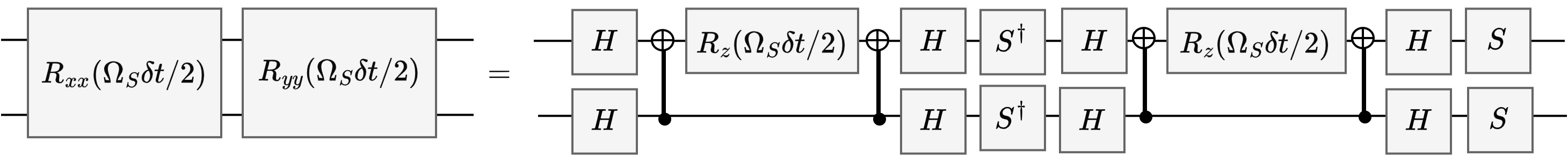} 
            \captionsetup{justification=centering}
            \caption{
            Quantum-circuit realization of the $S$-pulse.
            The circuit implements the combined two-qubit rotation 
            $R_{XX}(\Omega_S \delta t), \hspace{0.2cm}R_{YY}(\Omega_S \delta t)$ 
            through a sequence of native gates.
            Hadamard ($H$) and phase ($S, S^{\dagger}$) operations 
            transform the interaction basis, enabling experimental implementation 
            of the ideal $R_{XX}$ and $R_{YY}$ couplings that define the $S$-pulse evolution.
            }
            \label{fig:appB_Spulse}
        \end{minipage}%
    }
\end{figure}
\[
X=\ket{0}\!\bra{1}+\ket{1}\!\bra{0}, \qquad
Y=i\bigl(\ket{1}\!\bra{0}-\ket{0}\!\bra{1}\bigr)
\;\;\Longrightarrow\;\;
\ket{0}\!\bra{1}=\tfrac{X+iY}{2},\quad
\ket{1}\!\bra{0}=\tfrac{X-iY}{2}.
\]

\[
\begin{aligned}
\ket{0}\!\bra{1}\otimes\ket{0}\!\bra{1}
+\ket{1}\!\bra{0}\otimes\ket{1}\!\bra{0}
&=\left(\tfrac{X+iY}{2}\otimes\tfrac{X+iY}{2}\right)
 +\left(\tfrac{X-iY}{2}\otimes\tfrac{X-iY}{2}\right)\\
&=\frac{1}{4}\Big[(X\!\otimes\! X+iX\!\otimes\! Y+iY\!\otimes\! X-Y\!\otimes\! Y)\\
&\qquad\;\;\;\;+(X\!\otimes\! X-iX\!\otimes\! Y-iY\!\otimes\! X-Y\!\otimes\! Y)\Big]\\
&=\frac{1}{2}\bigl(X\!\otimes\! X - Y\!\otimes\! Y\bigr)
\end{aligned}
\]

\[
\ket{00}\!\bra{11}+\ket{11}\!\bra{00}=\frac{1}{2}\bigl(X\!\otimes\! X - Y\!\otimes\! Y\bigr)
\]

\[
\,H_P(t)=\frac{\Omega_P(t)}{4}\,\bigl(X\!\otimes\! X - Y\!\otimes\! Y\bigr)\, 
\]

For $S-$pulse, the Hamiltonian $H_{S}(t)$ is 

\begin{equation}
H_{S}(t)=\frac{\Omega_{S}(t)}{2}\left(|11\rangle\langle10|+|10\rangle\langle11|\right)\label{eqn 21}
\end{equation}

Using identity $|11\rangle\langle10|+|10\rangle\langle11|=2(\sigma_{0}^{+}\sigma_{1}^{-}+\sigma_{0}^{-}\sigma_{1}^{+})=\frac{1}{2}(X\otimes X+Y\otimes Y)$.
Putting the value of identity in Eq. (\ref{eqn 21})

\[
H_{S}(t)=\frac{\Omega_{S}(t)}{4}\left(X\otimes X+Y\otimes Y\right)
\]

The time evolution over a time interval $\delta t$ become

\[
U_{S}^{(i)}=e^{-i(H_{S}(t_{i})\delta t)}=e^{-i\left[\frac{\Omega_{S}(t_{i})}{4}\left(X\otimes X+Y\otimes Y\right)\right]\delta t}
\]

we can further define $XX$and $YY$ rotational gates with $\theta_{i}=\frac{\Omega_{S}(t_{i})}{4}\delta t$

\begin{equation}
U_{S}^{(i)}=e^{-i\left[\frac{\Omega_{S}(t_{i})}{4}\left(X\otimes X\right)\right]\delta t}.e^{-i\left[\frac{\Omega_{S}(t_{i})}{4}\left(Y\otimes Y\right)\right]\delta t}\label{eqn 22}
\end{equation}

The quantum circuit for $H_P(t)$ and $H_S(t)$ is shown in Fig. \ref{fig:appB_Ppulse} and  Fig. \ref{fig:appB_Spulse}. Crucially, the operators commute: $[X\otimes X, Y\otimes Y] = 0$. This commutation allows the time-evolution operators for both the P-pulse ($X\otimes X - Y\otimes Y$) and S-pulse ($X\otimes X + Y\otimes Y$) to be decomposed exactly into sequential rotations (e.g., $e^{-i H_S \delta t} = e^{-i \frac{\Omega_S}{4} X\otimes X \delta t} e^{-i \frac{\Omega_S}{4} Y\otimes Y \delta t}$) without introducing internal Trotter errors. This exact decomposition is vital for chiral discrimination, as it minimizes phase distortions that could otherwise disrupt the destructive interference required to maintain the dark state trajectory.

\setcounter{equation}{0}
\makeatletter
\renewcommand{\theequation}{C\arabic{equation}}
\makeatother

\refstepcounter{section}  
\section*{APPENDIX \Alph{section}: Derivation of Equation}
\phantomsection           
\label{app:C}

\begin{equation}
U_{0}(t)=\sum_{n=0,\pm}|\gamma_{n}\rangle\langle\gamma_{n}(t)|\label{eqn 33}
\end{equation}

\begin{equation}
H_{I}(t)=U_{0}(t)H_{PS}(t)U_{0}^{\dagger}(t)-iU_{0}(t)\dot{U}_{0}^{\dagger}(t)\label{eqn 34}
\end{equation}

(i)$\hspace{1em}U_{0}(t)H_{PS}(t)U_{0}^{\dagger}(t):$
\begin{align}
U_{0}(t) H_{PS}(t) U_{0}^{\dagger}(t) = &
\left[ \sum_{n=0,\pm} |\gamma_{n}\rangle \langle \gamma_{n}(t)| \right]
\left[ \Omega_{P}(t) |00\rangle\langle01| + \Omega_{S}(t) |01\rangle\langle10| \right] \nonumber \\
& \times \left[ \sum_{n=0,\pm} |\gamma_{n}(t)\rangle \langle \gamma_{n}| \right]
- i \left[ \sum_{n=0,\pm} |\gamma_{n}\rangle \langle \gamma_{n}(t)|
\left( \sum_{m=0,\pm} \frac{\partial}{\partial t} |\gamma_{m}(t)\rangle \langle \gamma_{m}| \right) \right]
\label{eqn:PS}
\end{align}

\begin{align}
U_{0}(t) H_{PS}(t) U_{0}^{\dagger}(t) =&\ \big[\,|\gamma_{+}\rangle\langle\gamma_{+}(t)|
      + |\gamma_{-}\rangle\langle\gamma_{-}(t)|\,\big] \nonumber \\
&\times \big[\,\Omega_{P}(t)|00\rangle\langle01|
      + \Omega_{S}(t)|01\rangle\langle10|
      + \text{h.c.}\,\big]
      \big[\,|\gamma_{+}(t)\rangle\langle\gamma_{+}|
      + |\gamma_{-}(t)\rangle\langle\gamma_{-}|\,\big]
\label{eqn:PS2}
\end{align}

\[
\mathrm{U}_{0}(t) H_{PS}(t) U_{0}^{\dagger}(t)
=\frac{\Omega_{P}^{2}(t)}{2\Omega}\,|\gamma_{+}\rangle\langle\psi_{+}|
+\frac{\Omega_{S}^{2}(t)}{2\Omega}\,|\gamma_{-}\rangle\langle\gamma_{-}|
\]

\begin{equation}
U_{0}(t)H_{PS}(t)U_{0}^{\dagger}(t)
=\Omega(t)\left[|\gamma_{+}\rangle\langle\gamma_{+}|
-|\gamma_{-}\rangle\langle\gamma_{-}|\right]
\label{eqn:diag2}
\end{equation}

For nonadiabatic coupling

\begin{equation}
-iU_{0}(t)\dot{U}_{0}^{\dagger}(t)=-i\left[|\gamma_{+}\rangle\langle\gamma_{+}(t)|\frac{\partial}{\partial t}|\gamma_{+}(t)\rangle\langle\gamma_{+}|
+|\gamma_{-}\rangle\langle\gamma_{-}(t)|\frac{\partial}{\partial t}|\gamma_{-}(t)\rangle\langle\gamma_{-}|\right]
\label{eqn:gamma-deriv}
\end{equation}
 
\begin{equation*}
\frac{\partial}{\partial t}|\gamma_{0}(t)\rangle
=-\dot{\alpha}_{1}\sin\alpha_{1}(t)|00\rangle
+\dot{\alpha}_{1}\cos\alpha_{1}(t)|10\rangle
\label{eqn:gamma0-deriv}
\end{equation*}

\vspace{0.5em}
\[
\implies
\]
\vspace{0.5em}

\begin{equation}
-iU_{0}(t)\dot{U}_{0}^{\dagger}(t)
=\frac{\dot{\alpha}_{1}(t)}{\sqrt{2}}\left[|\gamma_{+}\rangle\langle\gamma_{0}|
+|\psi_{-}\rangle\langle\gamma_{0}|\right]
\label{eqn:U0-deriv}
\end{equation}

The equation becomes

\begin{equation}
H_{I}(t) = \Omega(t)\big[\,|\gamma_{+}\rangle\langle\gamma_{+}| - |\gamma_{-}\rangle\langle\gamma_{-}|\,\big]
+ \frac{\dot{\alpha}_{1}(t)}{\sqrt{2}}\big[\,|\gamma_{+}\rangle\langle\gamma_{0}| + |\gamma_{-}\rangle\langle\gamma_{0}|\,\big]
+ \text{h.c.}
\label{eqn:35}
\end{equation}

\setcounter{equation}{0}
\makeatletter
\renewcommand{\theequation}{D\arabic{equation}}
\makeatother

\refstepcounter{section}  
\section*{APPENDIX \Alph{section}: Counterdiabatic Frame}
\phantomsection           
\label{app:D}

The goal of STAP is to cancel nonadiabatic transitions by
introducing a counteradiabatic auxiliary field $\Omega'_{P}(t)$ and
$\Omega'_{S}(t)$, and the Hamiltonian takes the form
\begin{equation}
H'(t)
= \big[\Omega_{P}(t)+\Omega'_{P}(t)\big]\ket{00}\bra{01}
+ \big[\Omega_{S}(t)+\Omega'_{S}(t)\big]\ket{01}\bra{10}
+ \text{h.c.}
\label{eqn:36}
\end{equation}

By applying \( U_{1}(t)=\sum_{n=0,\pm}\ket{\varphi_{n}}\!\bra{\varphi_{n}(t)} \)
to \( H'(t) \) the system is expressed in the adiabatic frame

\begin{equation}
H'_{\text{AD}}(t) = U_{1}(t) H'(t) U_{1}^{\dagger}(t) - i U_{1}(t) \dot{U}_{1}^{\dagger}(t)
\end{equation}

(i)\;$U_{1}(t)H^{'}(t)U_{1}^{\dagger}(t):$

\begin{equation}
U_{1}(t) H'(t) U_{1}^{\dagger}(t) 
= \left[ \sum_{n=0,\pm} |\varphi_{n}\rangle \langle \varphi_{n}(t)| \right]
\Big[ (\Omega_{P}(t) + \Omega_{P}'(t)) |00\rangle \langle01|
+ (\Omega_{S}(t) + \Omega_{S}'(t)) |01\rangle \langle10|
+ \text{h.c.} \Big]
\left[ \sum_{n=0,\pm} |\varphi_{n}(t)\rangle \langle \varphi_{n}| \right]
\label{eq:adiabatic-transformation}
\end{equation}

\begin{align}
U_{1}(t) H'(t) U_{1}^{\dagger}(t) = &\ \Big[ |\varphi_{+}\rangle \langle \varphi_{+}(t)|
+ |\varphi_{-}\rangle \langle \varphi_{-}(t)| \Big] \nonumber \\
&\times \Big[ (\Omega_{P}(t) + \Omega_{P}'(t)) |00\rangle \langle01|
+ (\Omega_{S}(t) + \Omega_{S}'(t)) |01\rangle \langle10| \Big]
\Big[ |\varphi_{+}(t)\rangle \langle \varphi_{+}|
+ |\varphi_{-}(t)\rangle \langle \varphi_{-}| \Big]
\label{eqn:XX}
\end{align}

\[
=\frac{1}{2}(\Omega_{P}(t)+\Omega_{P}'(t))[-\sin\alpha_{1}(t)\cos\alpha_{2}(t)-i\sin\alpha_{2}(t)\cos\alpha_{1}(t)\cos\alpha_{2}(t)]|\varphi_{+}\rangle\langle\varphi_{+}|+
\]

\[
\frac{1}{2}(\Omega_{P}(t)+\Omega_{P}'(t))[\sin\alpha_{1}(t)\cos\alpha_{2}(t)-i\sin\alpha_{2}(t)\cos\alpha_{1}(t)\cos\alpha_{2}(t)]|\varphi_{-}\rangle\langle\varphi_{-}|+
\]

\[
\frac{1}{2}(\Omega_{S}(t)+\Omega_{S}'(t))[\cos\alpha_{1}(t)\cos\alpha_{2}(t)+i\sin\alpha_{2}(t)\sin\alpha_{1}(t)\cos\alpha_{2}(t)]|\varphi_{+}\rangle\langle\varphi_{+}|+
\]

\[
\frac{1}{2}(\Omega_{S}(t)+\Omega_{S}'(t))[-\cos\alpha_{1}(t)\cos\alpha_{2}(t)-i\sin\alpha_{2}(t)\sin\alpha_{1}(t)\cos\alpha_{2}(t)]|\varphi_{-}\rangle\langle\varphi_{-}|
\]

Solving the real part first

\[
=\frac{1}{2}\left[\left\{ \left(\Omega_{P}(t)+\Omega_{P}'(t)\right)\left(-\sin\alpha_{1}(t)\right)\right\} +\left\{ \left(\Omega_{S}(t)+\Omega_{S}'(t)\right)\left(\cos\alpha_{1}(t)\right)\right\} \right]\cos\alpha_{2}|\varphi_{+}\rangle\langle\varphi_{+}|
\]

\[
+\frac{1}{2}\left[\left\{ \left(\Omega_{P}(t)+\Omega_{P}'(t)\right)\left(\sin\alpha_{1}(t)\right)\right\} +\left\{ \left(\Omega_{S}(t)+\Omega_{S}'(t)\right)\left(-\cos\alpha_{1}(t)\right)\right\} \right]\cos\alpha_{2}(t)|\varphi_{-}\rangle\langle\varphi_{-}|
\]

\[
U_{1}(t) H'(t) U_{1}^{\dagger}(t) =\varUpsilon(t)\left[|\varphi_{+}\rangle\langle\varphi_{+}|-|\varphi_{-}\rangle\langle\varphi_{-}|\right]
\]

where $\quad\varUpsilon(t)=\left[\Omega_{S}^{'}(t)\cos\alpha_{1}(t)-\Omega_{P}'(t)\sin\alpha_{1}(t)+\Omega\cos2\alpha_{1}(t)\right]\cos\alpha_{2}(t).$

Solving the imaginary part

\[
\frac{1}{2}\left[(\Omega_{P}(t)+\Omega_{P}'(t))(-\sin\alpha_{2}(t)\cos\alpha_{1}(t))+(\Omega_{S}(t)+\Omega_{S}'(t))(i\sin\alpha_{2}\sin\alpha_{1})\right](\cos\alpha_{2}(t))|\varphi_{+}\rangle\langle\varphi_{+}|
\]

\[
+\frac{1}{2}\left[(\Omega_{P}(t)+\Omega_{P}'(t))(-\sin\theta_{2}(t)\cos\alpha_{1}(t))+(\Omega_{S}(t)+\Omega_{S}'(t))(-i\sin\theta_{2}(t)\sin\alpha_{1}(t))\right](\cos\theta_{2}(t))|\varphi_{-}\rangle\langle\varphi_{-}|
\]

After including the h.c. terms, the contributions cancel, and the imaginary part vanishes. Thus, 
\( U_{1}(t)H'(t)U_{1}^{\dagger}(t) \) becomes

\[
U_{1}(t)H^{'}(t)U_{1}^{\dagger}(t)=\varUpsilon(t)\left[|\varphi_{+}\rangle\langle\varphi_{+}|-|\varphi_{-}\rangle\langle\varphi_{-}|\right]
\]

Now, we calculate $iU(t)\dot{U}^{\dagger}(t)$

\begin{equation}
\label{eq:UdotU}
iU(t)\dot{U}^{\dagger}(t)
=
\left[\sum_{n=0,\pm}|\varphi_{n}\rangle\langle\varphi_{n}(t)|\right]
\frac{\partial}{\partial t}
\left[\sum_{n=0,\pm}|\varphi_{n}(t)\rangle\langle\varphi_{n}|\right]
\end{equation}

\begin{equation}
\label{eq:UdotU_split}
iU(t)\dot{U}^{\dagger}(t)
=
\bigl[|\varphi_{+}(t)\rangle\langle\varphi_{+}(t)|\bigr]
\frac{\partial}{\partial t}
\bigl[|\varphi_{0}(t)\rangle\langle\varphi_{0}|\bigr]
+
\bigl[|\varphi_{-}(t)\rangle\langle\varphi_{-}(t)|\bigr]
\frac{\partial}{\partial t}
\bigl[|\varphi_{0}(t)\rangle\langle\varphi_{0}|\bigr]
\end{equation}

\[
\frac{\partial}{\partial t}|\varphi_{0}(t)\rangle=\dot{\alpha}_{2}\sin\alpha_{2}(t)\left[\cos\alpha_{1}(t)|00\rangle+\sin\alpha_{1}(t)\dot{\alpha}_{2}|10\rangle\right]+\cos\alpha_{2}(t)\left[-\dot{\alpha}_{1}(t)\sin\alpha_{1}(t)|00\rangle+\dot{\alpha}_{1}(t)\cos\alpha_{1}(t)|10\rangle\right]
\]

\[
+e^{i\phi}\dot{\alpha}_{2}\cos\alpha_{2}(t)|11\rangle
\]

\[
\langle\varphi_{+}(t)|\frac{\partial}{\partial t}|\varphi_{0}(t)\rangle=\frac{1}{\sqrt{2}}\left[\left\{ (\sin\alpha_{1}(t)+i\sin\alpha_{2}(t)\cos\alpha_{1}(t))\right\} \langle00|+\cos\alpha_{2}(t)\langle11|-\left\{ (\cos\alpha_{1}(t)+i\sin\alpha_{2}(t)\sin\alpha_{1}(t))\right\} \langle10|\right]
\]

\[
\left[-\dot{\alpha}_{2}\sin\alpha_{2}(t)\left\{ (\cos\alpha_{1}(t)|00\rangle+\sin\alpha_{1}(t)|10\rangle)\right\} +\cos\alpha_{2}(t)\left\{ (-\dot{\alpha}_{1}\sin\alpha_{1}(t)|00\rangle+\dot{\alpha}_{1}\cos\alpha_{1}(t)|10\rangle)\right\} +e^{i\phi}\dot{\alpha}_{1}\cos\alpha_{2}(t)|11\rangle\right]
\]

\[
=\frac{1}{\sqrt{2}}\left[\left\{ \sin\alpha_{1}+i\sin\alpha_{2}\cos\alpha_{1}\right\} \left\{ -\dot{\alpha}_{2}\sin\alpha_{2}\cos\alpha_{1}\right\} +\left\{ \sin\alpha_{1}+i\sin\alpha_{2}\cos\alpha_{1}\right\} \left\{ -\cos\alpha_{2}\dot{\alpha}_{1}\sin\alpha_{1}\right\} )-\cos^{2}\alpha_{2}e^{i\phi}\right.
\]

\[
\left.-\left\{ \cos\alpha_{1}-i\sin\alpha_{2}\sin\alpha_{1}\right\} \left\{ -\dot{\alpha}_{2}\sin\alpha_{2}\sin\alpha_{1}\right\} -\left\{ \cos\alpha_{1}-i\sin\alpha_{2}\sin\alpha_{1}\right\} \left\{ \cos\alpha_{2}\dot{\alpha}_{1}\cos\alpha_{1}\right\} \right]
\]

\begin{align*}
=\frac{1}{\sqrt{2}}\Bigl[&\,
   -\dot{\alpha}_{2}\sin\alpha_{1}\sin\alpha_{2}\cos\alpha_{1}
   - i\,\dot{\alpha}_{2}\sin^{2}\alpha_{2}\cos^{2}\alpha_{1}
   - \dot{\alpha}_{1}\sin^{2}\alpha_{1}\cos\alpha_{2}\\
&\quad
   - i\,\dot{\alpha}_{1}\sin\alpha_{1}\sin\alpha_{2}\cos\alpha_{1}\cos\alpha_{2}
   - e^{i\phi}\dot{\alpha}_{2}\cos^{2}\alpha_{2}
   + \dot{\alpha}_{2}\sin\alpha_{1}\sin\alpha_{2}\cos\alpha_{1}\\
&\quad
   - i\,\dot{\alpha}_{2}\sin^{2}\alpha_{2}\sin^{2}\alpha_{1}
   - \dot{\alpha}_{1}\cos^{2}\alpha_{1}\cos\alpha_{2}
   + i\,\dot{\alpha}_{1}\cos\alpha_{1}\cos\alpha_{2}\sin\alpha_{1}\sin\alpha_{2}
\Bigr].
\end{align*}

\[
=-\frac{1}{\sqrt{2}}\left[\dot{\alpha_{2}}\sin\alpha_{1}\sin\alpha_{2}\cos\alpha_{1}+\dot{\alpha_{1}}\sin^{2}\alpha_{1}\cos\alpha_{2}\right]-\frac{i}{\sqrt{2}}\left[\dot{\alpha_{2}}\sin^{2}\alpha_{2}\cos^{2}\alpha_{1}+\dot{\alpha_{1}}\sin\alpha_{1}\sin\alpha_{2}\cos\alpha_{1}\cos\alpha_{2}\right]
\]

\begin{multline*}
=\frac{i}{\sqrt{2}}\bigl[e^{i\phi}\dot{\alpha_{2}}\cos^{2}\alpha_{2}\bigr]
+\frac{1}{\sqrt{2}}\bigl[\dot{\alpha_{2}}\sin\alpha_{1}\sin\alpha_{2}\cos\alpha_{1}
-\dot{\alpha_{1}}\cos^{2}\alpha_{1}\cos\alpha_{2}\bigr]\\
{}+\frac{i}{\sqrt{2}}\bigl[\dot{\alpha_{2}}\sin^{2}\alpha_{2}\sin^{2}\alpha_{1}
 +\dot{\alpha_{1}}\cos\alpha_{1}\cos\alpha_{2}\sin\alpha_{1}\sin\alpha_{2}\bigr]
\end{multline*}

Solving the real part first 
\[
=-\frac{1}{\sqrt{2}}\left[\dot{\alpha_{2}}\sin\alpha_{1}\sin\alpha_{2}\cos\alpha_{1}+\dot{\alpha_{1}}\sin^{2}\alpha_{1}\cos\alpha_{2}\right]+\frac{1}{\sqrt{2}}\left[\dot{\alpha_{2}}\sin\alpha_{1}\sin\alpha_{2}\cos\alpha_{1}-\dot{\alpha_{1}}\cos^{2}\alpha_{1}\cos\alpha_{2}\right]
\]

\[
=\frac{1}{\sqrt{2}}\dot{\alpha}_{1}\cos\alpha_{2}\left[\sin^{2}\alpha_{1}+\cos^{2}\alpha_{1}\right]
\]

\begin{equation}
=\frac{1}{\sqrt{2}}\dot{\alpha}_{1}\cos\alpha_{2}
\label{eqn:alpha1}
\end{equation}

Solving the imaginary part

\begin{align*}
&=-\frac{i}{\sqrt{2}}\Bigl[\dot{\alpha_{2}}\sin^{2}\alpha_{2}\cos^{2}\alpha_{1}
   +\dot{\alpha_{1}}\sin\alpha_{1}\sin\alpha_{2}\cos\alpha_{1}\cos\alpha_{2}\Bigr] \\
&\quad -\frac{i}{\sqrt{2}}\Bigl[e^{i\phi}\dot{\alpha_{2}}\cos^{2}\alpha_{2}\Bigr]
+\frac{i}{\sqrt{2}}\Bigl[\ddot{\alpha_{2}}\sin^{2}\alpha_{2}\sin^{2}\alpha_{1}
   +\dot{\alpha_{1}}\cos\alpha_{1}\cos\alpha_{2}\sin\alpha_{1}\sin\alpha_{2}\Bigr]
\end{align*}

\[
=\frac{i}{\sqrt{2}}\left[-\dot{\alpha}_{2}\sin^{2}\alpha_{2}\left(\cos^{2}\alpha_{1}+\sin^{2}\alpha_{1}\right)-\dot{\alpha}_{2}\cos^{2}\alpha_{2}+2\dot{\alpha}_{1}\sin\alpha_{2}\sin\alpha_{1}\cos\alpha_{1}\cos\alpha_{2}\right]
\]

\[
=\frac{i}{\sqrt{2}}\left[+\dot{\alpha}\sin^{2}\alpha_{2}\cos2\alpha_{1}-\dot{\alpha}_{2}\cos^{2}\alpha_{2}+\dot{\alpha}_{1}\sin\alpha_{2}\sin2\alpha_{1}\cos\alpha_{2}\right]
\]

$\implies$

\[
-i\langle\varphi_{+}(t)|\dot{\varphi}_{0}(t)\rangle=\frac{1}{\sqrt{2}}\left[i\dot{\alpha}_{1}\cos\alpha_{2}+\dot{\alpha}_{2}\sin^{2}\alpha_{2}\cos2\alpha_{1}-\dot{\alpha}_{2}\cos^{2}\alpha_{2}+\dot{\alpha}_{1}\sin\alpha_{1}\sin2\alpha_{2}\cos\alpha_{2}\right]
\]

\begin{equation}
-i\langle\varphi_{+}(t)|\dot{\varphi}_{0}(t)\rangle
=\frac{1}{\sqrt{2}}\left[\dot{\alpha}_{1}\left\{ i\cos\alpha_{2}
+\sin\alpha_{1}\sin2\alpha_{2}\cos\alpha_{2}\right\}
+\dot{\theta}_{2}\left\{ \sin^{2}\alpha_{2}\cos2\alpha_{1}
-\cos^{2}\alpha_{2}\right\} \right]
\label{eqn:phi-plus}
\end{equation}

Combining Eq.(\ref{eqn:alpha1}) and Eq. (\ref{eqn:phi-plus}), we get

\begin{equation}
\label{eq:UdotU_lambda}
iU(t)\dot{U}^{\dagger}(t)
=
\bigl[\lambda_{+}(t)\ket{\varphi_{+}(t)}\bra{\varphi_{0}(t)}
+\lambda_{-}(t)\ket{\varphi_{-}(t)}\bra{\varphi_{0}(t)}
+\mathrm{h.c.}\bigr]
\end{equation}

where 

\begin{align*}
\lambda_{\pm}(t) ={}&\,i\Bigl\{\dot{\alpha}_1(t)\cos\alpha_2(t)
   - \sin\theta_2(t)\bigl[\Omega'_S(t)\cos\alpha_1(t)
   - \Omega'_P(t)\sin\alpha_1(t)
   + \Omega\cos2\alpha_1(t)\bigr]\Bigr\}\\
&\quad\mp\bigl[\Omega'_P(t)\cos\alpha_1(t)
   - \Omega'_S(t)\sin\alpha_1(t)
   + \Omega\sin2\alpha_1(t)
   + \dot{\alpha}_2(t)\bigr]
\end{align*}

\setcounter{equation}{0}
\makeatletter
\renewcommand{\theequation}{E\arabic{equation}}
\makeatother

\refstepcounter{section}  
\section*{APPENDIX \Alph{section}: Amplitude Noise and Infidelity Derivation}
\phantomsection           
\label{app:E}

\text{Consider small fractional amplitude noise:}
\[
\Omega_P \rightarrow \Omega_P (1 + \delta_P), \quad 
\Omega_S \rightarrow \Omega_S (1 + \delta_S),
\]
where $\abs{\delta_{P,S}} \ll 1$

Define
\[
r = \frac{\Omega_P}{\Omega_S}.
\]
Then
\[
r \rightarrow r' = \frac{\Omega_P(1+\delta_P)}{\Omega_S(1+\delta_S)}
= r \frac{1+\delta_P}{1+\delta_S}
\]

We simplify $\dfrac{1+\delta_P}{1+\delta_S}$ using first-order approximation:
\[
f(\delta_S,\delta_P) = \frac{1+\delta_P}{1+\delta_S} \approx 1 + \delta_P - \delta_S
\]

As the mixing angle is
\[
\alpha_1 = \arctan(r),
\]
the change in $\alpha_1$ due to small variation in $r$ is
\[
\delta \alpha_1 = \alpha_1(r + \delta r) - \alpha_1(r)
\]

\text{Using Taylor expansion}
\[
\delta \alpha_1 = \frac{d\alpha_1}{dr}\, \delta r
\]

Since $\alpha_1 = \arctan(r)$,
\[
\frac{d\alpha_1}{dr} = \frac{1}{1+r^2}
\]

Therefore,
\[
\delta \alpha_1 = \frac{1}{1+r^2}\, \delta r
\]

Using $\delta r = r (\delta_P - \delta_S)$, we get
\[
\boxed{
\delta \alpha_1 = \frac{r}{1+r^2} (\delta_P - \delta_S)
= \frac{\Omega_P \Omega_S}{\Omega_P^2 + \Omega_S^2} (\delta_P - \delta_S)
}
\]

\bigskip
\text{From ideal dark state}
\[
\ket{\gamma_0} = \cos(\alpha_1)\ket{00} - \sin(\alpha_1)\ket{10}
\]

Due to amplitude noise, the mixing angle is perturbed
\[
\alpha_1 \rightarrow \alpha_1 + \delta\alpha_1
\]
Hence the new, perturbed dark state is
\[
\ket{\tilde{\gamma}_0} = \cos(\alpha_1 + \delta\alpha_1)\ket{00}
- \sin(\alpha_1 + \delta\alpha_1)\ket{10}
\]

For small $\delta\alpha_1$,
\[
\cos(\delta\alpha_1) \approx 1, \quad \sin(\delta\alpha_1) \approx \delta\alpha_1
\]

Therefore,
\begin{align*}
\ket{\tilde{\gamma}_0}
&= [\cos\alpha_1 - \delta\alpha_1 \sin\alpha_1]\ket{00}
- [\sin\alpha_1 + \delta\alpha_1 \cos\alpha_1]\ket{10} \\
&= [\cos\alpha_1\ket{00} - \sin\alpha_1\ket{10}]
- \delta\alpha_1[\sin\alpha_1\ket{00} + \cos\alpha_1\ket{10}]
\end{align*}

At final time $T$, $\alpha_1(T) = \pi/2$, so
\[
\ket{\tilde{\gamma}_0(T)} = -\ket{10} - \delta\alpha_1(T)\ket{00}
\]

\text{The goal} is to transfer population to $\ket{10}$

\bigskip
\text{Infidelity definition}
\[
1 - F = |\braket{\text{target} | \text{actual}}|^2
\]
Target state: $\ket{10}$.

\begin{align*}
\braket{\text{target} | \text{actual}}
&= \braket{10 | (-\ket{10} - \delta\alpha_1(T)\ket{00})} \\
&= -1 - 0 = -1
\end{align*}

Hence the infidelity arises from the leakage term:
\[
1 - F = |\braket{00 | \tilde{\gamma}_0(T)}|^2 = |\delta\alpha_1(T)|^2
\]

\[
\boxed{1 - F = |\delta\alpha_1(T)|^2}
\]

\end{document}